\documentstyle[aaspp4,bsf,11pt]{article}

\slugcomment{accepted for publication in ApJ}

\begin{document}

\title
{Magnetorotational Instability in Protoplanetary Disks. \\
II. Ionization State and Unstable Regions}
\author
{Takayoshi Sano\altaffilmark{1}, Shoken M. Miyama} 
\affil
{National Astronomical Observatory, Mitaka, Tokyo 181-8588, Japan}
\authoraddr
{Department of Astronomy, University of Maryland, College Park, MD
20742-2421}
\authoremail
{sano@astro.umd.edu}

\author
{Toyoharu Umebayashi}
\affil
{Computing Service Center, Yamagata University, Yamagata 990-8560,
Japan}

\and

\author
{Takenori Nakano}
\affil
{Department of Physics, Kyoto University, Sakyo-ku, Kyoto 606-8502,
Japan}

\altaffiltext{1}{present address: Department of Astronomy, University of
Maryland, College Park, MD 20742-2421; sano@astro.umd.edu}

\begin{abstract}
We investigate where in protoplanetary disks magnetorotational
 instability operates, which can cause angular momentum transport in the
 disks.
We investigate the spatial distribution of various charged particles and
 the unstable regions for a variety of models for protoplanetary disks
 taking into account the recombination of ions and electrons at grain
 surfaces, which is an important process in most parts of the disks. 
We find that for all the models there is an inner region which is
 magnetorotationally stable due to ohmic dissipation. 
This must make the accretion onto the central star non-steady.
For the model of the minimum-mass solar nebula, the critical radius,
 inside of which the disk is stable, is about 20 AU, and the mass
 accretion rate just outside the critical radius is $10^{-7}$ --
 $10^{-6} ~M_{\odot}~{\rm yr}^{-1}$.
The stable region is smaller in a disk of lower column density.
Dust grains in protoplanetary disks may grow by mutual sticking and may
 sediment toward the midplane of the disks.
We find that the stable region shrinks as the grain size increases or
 the sedimentation proceeds.
Therefore in the late evolutionary stages, protoplanetary disks can be
 magnetorotationally unstable even in the inner regions.
\end{abstract}


\section{INTRODUCTION} 

Various infrared and radio observations have revealed the existence of
accretion disks around T Tauri stars, so-called protoplanetary disks,
where planet formation may be going on (e.g., Rucinski 1985; Adams,
Lada, \& Shu 1987). 
Photometric observations at optically thin millimeter wavelengths give
disk masses of $\sim 0.1$ -- $0.01 ~M_{\odot}$ (Beckwith et al. 1990),
and fitting of spectral energy distributions (SEDs) at infrared
wavelengths gives disk sizes of tens of AU or 100 AU (Adams, Lada, \&
Shu 1988). 
Some authors tried to reproduce the SEDs of T Tauri stars with models
of viscous accretion disks (e.g., Bertout, Basri, \& Bouvier 1988), and
succeeded in deriving the slopes of SEDs at infrared wavelengths for
some T Tauri stars.
Optical and ultraviolet veiling of absorption lines has been discovered
for some T Tauri stars, which indicates the existence of boundary layers
at the interface between a Keplerian accretion disk and a slowly
spinning star (Lynden-Bell \& Pringle 1974). 
All these suggest the existence of accreting matter in protoplanetary
disks at least in their early active stages. 
However, the mechanisms of angular momentum transport, necessary for
accretion, have not yet been well understood, any more than in other
astrophysical accretion disks.  

The magnetorotational instability must be the most promising source of
anomalous viscosity in accretion disks (Balbus \& Hawley 1998, and
references therein).
Balbus \& Hawley (1991) showed that astrophysical accretion disks are
unstable to axisymmetric disturbances in the presence of a weak magnetic
field.
The instability sets in under a broad range of conditions and is capable
of initiating and sustaining MHD turbulence in accretion disks, as long
as the approximation of the ideal MHD holds.
However, protoplanetary disks are so cold and dense that the ionization
fraction is very low (Umebayashi \& Nakano 1988). 
Because the coupling of some charged particles with magnetic fields is
not strong enough, both the processes of ambipolar diffusion and ohmic
dissipation have to be considered (Nakano 1984). 
These processes have an effect of suppressing the magnetorotational
instability.
It is necessary to clarify which parts of the disks are
magnetorotationally unstable in relation to the evolution of
protoplanetary disks. 

Blaes \& Balbus (1994) examined the effect of the ambipolar diffusion on
the instability, and showed that differentially rotating disks are
unstable when the collision frequency of neutral particles with ions
exceeds the local epicyclic frequency. 
Sano \& Miyama (1999, hereafter Paper I) investigated the stability of
weakly ionized disks including the effect of ohmic dissipation. 
This process dominates the ambipolar diffusion in the regions of higher
density, or in the inner parts (within about 100 AU) of the disks.

In paper I, we derived the conditions for the existence of unstable
modes by the global linear analysis taking into account the effect of
the vertical structure of the disks.
The wavelength of the most unstable mode expected from the local linear
analysis is approximately given by
\begin{equation}
\lambda_{\rm local}
\approx \max 
\left( 
2 \pi \frac{v_{\rm A}}{\Omega}, 
2 \pi \frac{\eta}{v_{\rm A}}
\right) \;,
\label{eqn:lambda}
\end{equation}
where $v_{\rm A} = B / ( 4 \pi \rho )^{1/2}$ is the Alfv{\'e}n speed,
$\Omega$ is the angular rotational velocity of the disk, and $\eta$ is
the magnetic diffusivity (Jin 1996; Paper I).
If the magnetic Reynolds number $R_{\rm m} \equiv v_{\rm A}^2 / \eta
\Omega$ is larger than unity, the ohmic dissipation is not efficient
and the wavelength of the most unstable mode is approximately given by
$2 \pi v_{\rm A} / \Omega$, which is the result of the ideal MHD.
When $R_{\rm m} \lesssim 1$, on the other hand, perturbations with
shorter wavelengths are stabilized due to the dissipation.
Thus, as the magnetic diffusivity increases, $2 \pi \eta / v_{\rm A}$
increases and finally becomes the wavelength of the most unstable mode.

The local analysis gives the wavelength $\lambda_{\rm local}$ of the
most unstable mode as a function of the local values of $v_{\rm A}$ and
$\eta$. 
Thus $\lambda_{\rm local}$ takes different values at different points in
the disk.
Our results of the global analysis show that the vertical distribution
of the wavelength $\lambda_{\rm local} (z)$ is quite important for the
understanding of the global stability.
In paper I, we have found that the layer at a height $z$ is globally
unstable when $\lambda_{\rm local} (z)$ obtained by the local analysis
is shorter than the scale height of the disk $H$, or
\begin{equation}
\lambda_{\rm local} (z) \approx 
\max \left[ 
2 \pi \frac{v_{{\rm A}}(z)}{\Omega},
2 \pi \frac{\eta(z)}{v_{{\rm A}}(z)} 
\right] \lesssim H \;.
\label{eqn:rzcrit}
\end{equation}  
The condition given by equation~(\ref{eqn:rzcrit}) agrees with the
idea that the stability criterion is approximately given by the
requirement that the shortest wavelength for local instability is
smaller than the scale height of the disk.

Thus the ratio $\lambda_{\rm local} / H$ is a good indicator for the
existence of global unstable modes.
This ratio is sensitive to the magnetic diffusivity, which is
determined by the physical quantities in protoplanetary disks such as
the density, the temperature, and the ionization fraction.
Umebayashi \& Nakano (1988) examined the radial distribution of
magnetic diffusivity on the midplane for the so-called minimum-mass
solar nebula proposed by Hayashi, Nakazawa, \& Nakagawa (1985).
However, to make use of equation~(\ref{eqn:rzcrit}) as the stability
criterion of protoplanetary disks, we have to know the vertical
distribution of magnetic diffusivity.
In this paper we investigate the two-dimensional distribution of
magnetic diffusivity for various disk models, and reveal where in
protoplanetary disks the magnetorotational instability is operative.

Because the recombination processes are more effective in denser
regions, the ionization fraction is lower at higher density.
Thus the inner dense region of the disks would be decoupled from
magnetic fields except for the innermost region where the thermal
ionization is efficient.
However, even in the inner dense region, the gas near the surface layer
of the disks may be well ionized because the density is low.
Gammie (1996) proposed a layered accretion model which has the decoupled
region (dead zone) sandwiched by the ionized active layers.
The ionization by X-rays from the central object would also make a
similar structure (Glassgold, Najita, \& Igea 1997).
In such situations, the column densities of the active layers are one of 
the most important quantities for the dynamical evolution of
protoplanetary disks.
We shall calculate the column densities of these layers and compare them 
with the previous works.

The recombination of ions and electrons on grain surface is the dominant
process in the dense region. 
Therefore the characters of dust grains affect significantly the
ionization fraction and the column density of the active layers. 
In the previous works grain surface reactions were not included because
they were concerned mainly in the late phases of the disks when dust
grains of the disks have already settled toward the midplane.
In this paper we investigate the effect of dust grains on the abundances
of charged particles and the contribution of charged grains to the
interaction of the gas with magnetic fields in protoplanetary disks, and
show where the magnetorotational instability can be operative.

The plan of this paper is as follows.  
In \S 2 we describe the disk models adopted in this paper and the
magnetic diffusivity of the gas composed of various kinds of charged
particles. 
We also describe the reaction scheme which determines the abundance of
charged particles in ionization-recombination equilibrium.
In \S 3 we show numerical results on the spatial distribution of charged
particles, the time scale of magnetic field dissipation, and the
unstable regions in the disks.
Because charged grains can be the dominant charged particles at least
in some parts of the protoplanetary disks, their size and abundance are
highly important in determining the magnetic diffusivity.
Dust grains may grow by mutual sticking. 
They will sediment onto the midplane if the disk is quiescent. 
The coupling between the gas and magnetic field varies as the disk
evolves in such ways.
In \S 3 we also examine the unstable regions in some model disks which
must correspond to different evolutionary stages.
In \S4 we investigate the column densities of active layers and discuss
the evolutionary scenario of protoplanetary disks taking account of the
magnetorotational instability.
We also discuss the validity of our assumptions and some related
problems in \S 4. 
Finally we summarize our results in \S 5.

\section{MODEL AND FORMULATION}

\subsection{Structure of Protoplanetary Disks}
\label{sec:model}

Although various models have been proposed for protoplanetary disks,
there seem to be no really acceptable models yet. 
In view of such a situation we will investigate on a variety of models
and find out some firm results.
The surface density $\Sigma (r)$ of the protoplanetary disks is one of
the most important quantities in constructing the disk models.  
We assume a power-law distribution
\begin{equation}
\Sigma (r) = \Sigma_0 \left
( \frac{r}{1 ~{\rm AU}} \right)^{-q} \;,
\label{eqn:sigr}
\end{equation}
where $r$ is the distance from the central star. 
Hayashi (1981) found that the surface density of the primitive solar
nebula, which is restored from the mass distribution in the present
solar system, is well represented by equation~(\ref{eqn:sigr}) with $q =
3 / 2$ and $\Sigma_0 \approx 1.7 \times 10^3 ~{\rm g}~{\rm cm}^{-2}$.
This model is thus called the minimum-mass solar nebula.  
We examine mainly for the models with $q = 3 / 2$ and $\Sigma_0 = 1.7
\times 10^3 f_{\Sigma} ~{\rm g}~{\rm cm}^{-2}$ in this paper, where
$f_{\Sigma}$ is an arbitrary parameter. 
Equation~(\ref{eqn:sigr}) in this case must break down at large $r$
because the total mass of the disk should not diverge.  
In this work we examine the regions of $r \lesssim$ 100 AU.
Because the power index $q$ could take different values (e.g., Cassen
1994), we will also investigate for the other values of $q$.

For the temperature of the disks we also assume a power-law distribution,
\begin{equation}
T (r) = T_0 \left( \frac{r}{1 ~{\rm AU}} \right)^{-p} \;.
\label{eqn:temp}
\end{equation}
Except for a very early accretion phase, the temperature of the disk
is determined by the energy balance between absorption of stellar
radiation and emission of thermal radiation.  
After dust grains in the disk have sunk toward the midplane to some
extent, most part of the disk is nearly transparent to the stellar
light and the thermal radiation.
The equilibrium temperature in such a situation is given by $p = 1 / 2$
and $T_0 = 280$ K for the luminosity of the central star $L_{\ast} = 1
~L_{\odot}$.
We adopt $p = 1 / 2$ and $T_0 = 280$ K even for the opaque regions near
the midplane because it may be regarded as being in a thermal bath of
this temperature.  
For simplicity we also adopt this temperature distribution even for the
early opaque stage before significant sedimentation of dust grains
occurs because the results are not very sensitive to the temperature as
long as it is below several hundred Kelvin (see \S \ref{sec:temp}). 
With $p = 1 / 2$ and $T_0 = 280$ K, the sound velocity is given by
\begin{equation}
c_{\rm s} (r) = \left( \frac{k T}{\mu m_{\rm H}} \right)^{1/2} = 9.9
\times 10^4 
\left( \frac{r}{1 ~{\rm AU}} \right)^{-1/4}
\left( \frac{\mu}{2.34} \right)^{-1/2} {\rm cm}~\sec^{-1} \;,
\label{eqn:csr}
\end{equation}
where $k$ is the Boltzmann constant, $\mu$ is the mean molecular
weight of the gas, and $m_{\rm H}$ is the mass of a hydrogen atom.

In a gravitationally stable disk, the gravity perpendicular to the
disk is mainly contributed by the central star. 
The balance of the stellar gravity and the pressure force in the
$z$-direction determines the gas density at height $z$ from the midplane
as
\begin{equation}
\rho (r,z) = \rho_c (r) \exp \left[ - \frac{z^2}{H^2(r)} \right] \;,
\end{equation}
where $\rho_{\rm c} (r)$ is the density at the midplane and 
\begin{equation}
H(r) = \frac{\sqrt{2} c_{\rm s}}{\Omega}
\label{eqn:hr}
\end{equation}
is the scale height of the disk.
For the Keplerian angular velocity $\Omega = ( G M_{\ast} / r^3
)^{1/2}$, where $G$ is the gravitational constant and $M_{\ast}$ is
the mass of the central star, the density at the midplane is given by 
\begin{equation}
\rho_{\rm c} (r) = 1.4 \times 10^{-9} f_{\Sigma} 
\left( \frac{r}{1~{\rm AU}} \right)^{-11/4} 
\left( \frac{M_{\ast}}{M_{\odot}} \right)^{1/2} 
\left( \frac{\mu}{2.34} \right)^{1/2} {\rm g}~{\rm cm}^{-3} \;,
\label{eqn:rhocr}
\end{equation}
for $\Sigma$ given by equation~(\ref{eqn:sigr}) with $q = 3/2$ and for
$c_{\rm s}$ given by equation~(\ref{eqn:csr}).

The local stability analysis shows that a disk rotating around a star
with the Keplerian velocity is gravitationally stable only when $\Sigma
< \Omega c_{\rm s} / \pi G$ (Goldreich \& Lynden-Bell 1965).
The disk given by equations~(\ref{eqn:sigr}) and (\ref{eqn:csr}) with $q
= 3 / 2$ is gravitationally stable when the inequality,
\begin{equation}
f_{\Sigma} < 17.5 
\left( \frac{r}{100 ~{\rm AU}} \right)^{-1/4} 
\left( \frac{M_{\ast}}{M_{\odot}} \right)^{1/2} 
\left( \frac{\mu}{2.34} \right)^{-1/2} \;,
\end{equation}
is fulfilled.
Since we are concerned with gravitationally stable disks of radius $r
\lesssim$ 100 AU, it is sufficient to consider the cases of $f_{\Sigma}
\lesssim 10$.

\subsection{Magnetic Diffusivity}

The magnetic diffusivity $\eta$ is given in terms of the electrical
conductivity $\sigma_{\rm c}$ and the light speed $c$ by
\begin{equation}
\eta = \frac{c^2}{4 \pi \sigma_{\rm c}} \;.
\label{eqn:eta}
\end{equation}
The electrical conductivity is given by 
\begin{equation}
\sigma_{\rm c} = \sum_{\nu} \frac{( e q_{\nu} )^2 \tau_{\nu}
n_{\nu}}{m_{\nu}} \;,
\label{eqn:cond}
\end{equation}
where $\tau_{\nu}$ is the viscous damping time of motion relative to
neutral particles of a charged particle $\nu$ with mass $m_{\nu}$ and
electric charge $e q_{\nu}$, and $n_{\nu}$ is the number density of
particle $\nu$.  

Because the mass of an ion, $m_{\rm i}$, is generally much greater than
the mass of a neutral molecule, $m_{\rm n}$, we have for an ion
\begin{equation}
\tau^{-1}_{\rm i} = n_{\rm n} \langle \sigma v \rangle_{\rm i}
\frac{m_{\rm n}}{m_{\rm i}}
\;,
\end{equation}
where $n_{\rm n}$ is the number density of neutrals and $\langle
\sigma v \rangle_{\rm i}$ is the rate coefficient for the collision
between ions and neutrals averaged over the distribution of the
relative velocity.
The momentum-transfer cross section of a singly charged ion due to
electrical polarization force is given by (Osterbrock 1961) 
\begin{equation}
\sigma = 2.41 \pi \left( \frac{\alpha e^2}{m v^2} \right)^{1/2} \;,
\label{eqn:cross}
\end{equation}
where $\alpha$ is the polarizability of a neutral molecule, $e$ is the
elementary electric charge, $m$ is the reduced mass, and $v$ is the
relative velocity. 
Almost all ions at such low temperatures are singly charged.
We have $\alpha = 7.9 \times 10^{-25}$ cm$^{3}$ for an H$_{2}$
molecule and $\alpha = 2.1 \times 10^{-25}$ cm$^{3}$ for a He atom.
By taking the mass of an ion $m_{\rm i} \approx 34 ~m_{\rm H}$, for
example, we have $\langle \sigma v \rangle_{{\rm i-H}_2} \approx 1.8
\times 10^{-9} ~{\rm cm}^{3}~{\rm s}^{-1}$, which is independent of
$v$ as long as $\sigma$ is larger than the geometrical cross section. 
Such a rate coefficient due to polarization force is called the
Langevin's rate coefficient.

For an electron we have
\begin{equation}
\tau^{-1}_{\rm e} = n_{\rm n} \langle \sigma v \rangle_{\rm e} \;.
\end{equation}
In the previous works, the Langevin's rate was also used for the rate
coefficient for the collision between electrons and neutrals.
However, the collision cross sections obtained by experiments at low
electron energies are much smaller than $\sigma$ given by
equation~(\ref{eqn:cross}) and is nearly equal to a geometrical cross
section.
In this paper, we use the cross sections recommended by M. Hayashi
(1981) who compiled the experimental results (see Appendix).

We neglect the contribution of electron-ion collisions to the electrical
conductivity in equation~(\ref{eqn:cond}).
The contribution of electron-ion collision is $\tau_{\rm e} / \tau_{\rm
e-i}$ times the contribution of electron-neutral collision, where
$\tau_{\rm e-i}$ is the collision time of an electron with ions, and is
inefficient unless $n_{\rm e} / n_{\rm n} > 10^{-8}$ and $10^{-4}$ at $T
= 10$ and $10^3$ K, respectively.

For a grain of any charge, we have
\begin{equation}
\tau^{-1}_{\rm g} = n_{\rm n} \langle \sigma v \rangle_{\rm g}
\frac{m_{\rm n}}{m_{\rm g}} \;, 
\end{equation}
where $m_{\rm g}$ is the mass of a grain.  Because the mean free path
of neutrals is much greater than the size of a grain, we have for a
spherical grain of radius $a$, 
\begin{equation}
\langle \sigma v \rangle_{\rm g} = \frac{4 \pi}3 a^2 c_{\rm th} \;,
\end{equation}
where $c_{\rm th} = ( 8 k T / \pi \mu m_{\rm H} )^{1/2}$ is the thermal 
velocity of a neutral molecule.

\subsection{Reaction Scheme}

We investigate numerically the densities of various kinds of charged
particles in order to calculate the magnetic diffusivity $\eta$ given by
equation~(\ref{eqn:eta}).  
The rate equation for a constituent X$_i$ is given by
\begin{equation}
\frac{d n  ({\rm X}_i)}{d t} = \sum_j \gamma_{ij} n( {\rm X}_j ) +
\sum_{j, k} \beta_{ijk} n( {\rm X}_j ) n( {\rm X}_k ) \;,
\label{eqn:rate1}
\end{equation}
where $n( {\rm X}_i)$ is the number density of X$_i$, $\gamma_{ij}$ is
the rate coefficient for forming X$_i$ by ionization of X$_j$, and
$\beta_{ijk}$ is the rate coefficient of a two-body reaction including
reactions on dust grains.  
Because the gas in protoplanetary disks can be regarded as being in
ionization-recombination equilibrium (see \S 4.6 below),
equation~(\ref{eqn:rate1}) is reduced to
\begin{equation}
\sum_j \frac{\gamma_{ij}}{n_{\rm H}} x( {\rm X}_j ) +
\sum_{j, k} \beta_{ijk} x( {\rm X}_j ) x( {\rm X}_k ) = 0 \;,
\label{eqn:rate2}
\end{equation}
where $x ({\rm X}_i) \equiv n ({\rm X}_i) / n_{\rm H}$. 
We solve equation~(\ref{eqn:rate2}) at each position of the disk to
obtain the spatial distribution of magnetic diffusivity.

We consider the regions of the disk where the temperature is less than
several hundred Kelvin, and thus the thermal ionization is inefficient
compared with the ionization by non-thermal particles (Umebayashi 1983;
Nakano \& Umebayashi 1988).
In such regions ions and electrons are produced mainly through
ionization by cosmic rays and radioactivity, and vanish through
various recombination processes.  
Since these situations are quite similar to those in very dense
interstellar clouds, we adopt a simplified reaction scheme for dense
clouds investigated by Umebayashi \& Nakano (1980, 1990) and Nishi,
Nakano, \& Umebayashi (1991) with some modifications.  
Among the modifications are updating the reaction rates and adding some
species of charged particles such as H$_2^{+}$ and grains of charge $\pm
3 e$.
We also take into account the electrical polarization of grains for
collisions with ions and electrons.  
In the following we describe the reaction scheme adopted in this
paper.

\subsubsection{Reactions in Gas Phase}

The elements to be considered in our reaction scheme are H, He, C, O,
and refractory heavy elements.
We adopt the solar system abundances for these elements (Anders \&
Grevesse 1989) which are shown in Table~\ref{tbl:ele}.  
Most of the heavy elements have condensed in dust grains and only a
small fraction of them remain in the gas phase.  
We denote this fraction as $\delta_1$ for the volatile elements C and
O, and $\delta_2$ for the refractory heavy elements.
We assume for simplicity that all grains have the same size unless
otherwise stated.
Then the grain abundance $n_{\rm g} / n_{\rm H}$ by number relative to
hydrogen is determined from the total amounts of those elements in
grains.

We can regard that each element in gas phase has been transformed into
its molecular form because we consider low temperature, high density
regions.
We assume that all C in the gas phase is in CO molecule.
Oxygen in the gas phase other than the constituent of CO is in the
form of O$_2$, H$_2$O, OH, or O.  
Since the reaction rate coefficients of molecules O$_2$, H$_2$O, and
OH with ions are of the same order and the recombination rate
coefficients of the resultant ions with electrons are also of the same
order, we regard the reactions of the most abundant species O$_2$ as
the representatives of their reactions.  
We introduce a parameter $f_{{\rm O}_2}$ which represents the fraction
of oxygen in the gas phase in the molecular form O$_2$, H$_2$O, and
OH.  
Because the metal atoms such as Na, Mg, and Al have nearly the same rate
coefficients for charge-transfer reaction with molecular ions and the
resultant metal ions also have similar rate coefficients for radiative
recombination with electrons, we neglect the differences among them and
denote them as M collectively.

We consider the following species of charged particles in our reaction
scheme: electron e, H$^{+}$, He$^{+}$, C$^{+}$, H$_2^{+}$, H$_3^{+}$,
molecular ion m$^{+}$ (except H$_2^{+}$ and H$_3^{+}$), metal ion
M$^{+}$, and charged grains.  
Dust grains in the disks can be charged through the collision with
electrons and ions in the gas phase.
We consider the grains of charge 0 (neutral), $\pm 1e$, $\pm 2e$, and
$\pm 3e$.
Table~\ref{tbl:two} shows the reactions in gas phase and their rate
coefficients (Millar, Farquhar, \& Willacy 1997).  
The important reactions of molecular ions m$^{+}$ are dissociative
recombination with electrons and charge transfer with metal atoms.
Because the rate coefficients of these reactions are similar among
various molecular ions, we regard all m$^{+}$ as a single species.  
For the rate coefficients of m$^{+}$, we use those of the most
abundant ion, HCO$^{+}$.  
Similarly, we regard the rate coefficients of Mg$^{+}$ as the
representatives of those of metal ions M$^{+}$. 

\subsubsection{Reactions on Grain Surfaces}

Dust grains may gradually settle to a very thin layer about the
midplane reducing the grain abundance in the major part of the disk
(Nakagawa, Nakazawa, \& Hayashi 1981). 
We define a parameter $f_{\rm g}$ that represents the abundance of
dust grains relative to that in molecular clouds.
The depletion parameter $f_{\rm g}$ may be regarded as representing
the evolutionary stage of the disk.

For simplicity, we assume that all dust grains have the same size
(radius) $a$, and that they consist of rocky and metallic materials.
We take the inner density of grain material as 3 g cm$^{-3}$.  
We discuss the effects of the grain-size distribution and the effect
of tiny grains in \S 3.3.  
We neglect icy material (H$_2$, CH$_4$, and NH$_3$) in dust grains
because they have only small effects on the ionization state of the
disk (Umebayashi 1983).

At the low temperatures concerned in this work, dust grains can adsorb
ions and electrons.
Thus, in addition to the radiative and dissociative recombinations of
ions and free electrons in the gas phase, we have to take into account
their recombination on grain surfaces, which occurs through grain-ion,
grain-electron, and grain-grain collisions.  
As in the previous works (Nakano 1984; Umebayashi \& Nakano 1990;
Nishi et al. 1991), we adopt the following model for the reactions on
grain surfaces:
\begin{enumerate}
\item 
When an ion hits a neutral grain or a positively charged grain, it
sticks to the grain with a probability $S_{\rm i} \approx 1$.  
The sticking probability $S_{\rm e}$ of an electron onto a neutral or
negatively charged grain has been found to be between 0.2 and 1.0
(Nishi et al. 1991).  
We use $S_{\rm e} = 0.6$ as in the previous works.  
\item 
When an ion hits a negatively charged grain, or an electron hits a
positively charged grain, the electron and the ion recombine and leave
the grain surface using the energy released by the recombination.
\item 
When grains of opposite charge collide in thermal Brownian motion,
they neutralize themselves.  
We neglect coalescence of grains. 
\end{enumerate}

We take account of the effect of electrical polarization of grains
for the collision with an ion or an electron by using the fitting
formula for the collision rate coefficient obtained by Draine \& Sutin
(1987).

\subsubsection{Ionization Processes}

In protoplanetary disks, charged particles are formed first by
ionization of H$_2$ and He by cosmic rays and radioactive elements.
Table~\ref{tbl:ionize} lists the reaction rates of H$_2$ and He with
such energetic particles in units of the total ionization rate $\zeta$
of a hydrogen molecule, which includes the ionization by secondary
electrons.  
The ionization rate in the disk is effectively given by
\begin{equation}
\zeta (r,z) \approx \frac{\zeta_{\rm CR}}{2} 
\left[ \exp \left( -
\frac{\chi (r,z)}{\chi_{\rm CR}} \right) 
+ \exp \left( - \frac{\Sigma
(r) - \chi (r,z)}{\chi_{\rm CR}} \right) \right] 
+ \zeta_{\rm R} \;,
\label{eqn:zetaz}
\end{equation}
where $\zeta_{\rm CR} \approx 1.0 \times 10^{-17} \sec^{-1}$ is the
ionization rate by cosmic rays in the interstellar space, and
$\zeta_{\rm R} \approx 6.9 \times 10^{-23} f_{\rm g}~\sec^{-1}$ is the 
ionization rate by radioactive elements contained in the disk
(Umebayashi \& Nakano 1981).
The depth
\begin{equation}
\chi (r,z) = \int^{\infty}_{z} \rho (r,z) dz 
\end{equation}
is the vertical column density from infinity to the position concerned.
The attenuation length of the ionization rate by cosmic rays has a
value $\chi_{\rm CR} \approx 96$ ${\rm g}~{\rm cm}^{-2}$ (Umebayashi
\& Nakano 1981).
Because most of the radioactive elements are in rocky and metallic
materials, the rate $\zeta_{\rm R}$ is proportional to the depletion
factor $f_{\rm g}$ of dust grains. 

\section{RESULTS}

\subsection{The Minimum-Mass Solar Nebula --- The Fiducial Model}

First, we investigate the stability of a disk rotating around a star
of $M_{\ast} = 1 ~M_{\odot}$ with the surface density parameter
$f_{\Sigma} = 1$, which we presume to well represent the primitive solar
nebula.  
We take $p = 1 / 2$ and $T_0 = 280$ K in equation~(\ref{eqn:temp}).
We adopt the radius of dust grains $a = 0.1$ $\mu$m and the grain
depletion factor $f_{\rm g} = 1$ corresponding to early evolutionary
stages where sedimentation of dust grains has hardly proceeded yet.
We assume $\delta_1 = 0.2$ and $\delta_2 = 0.02$ for the parameters of
element depletion, and we adopt $f_{{\rm O}_2} = 0.7$ in accordance 
with the results of Mitchell, Kuntz, \& Ginsburg (1978) on the
calculation of molecular abundances.
We call this the fiducial model.

\subsubsection{Spatial Distribution of Charged Particles}

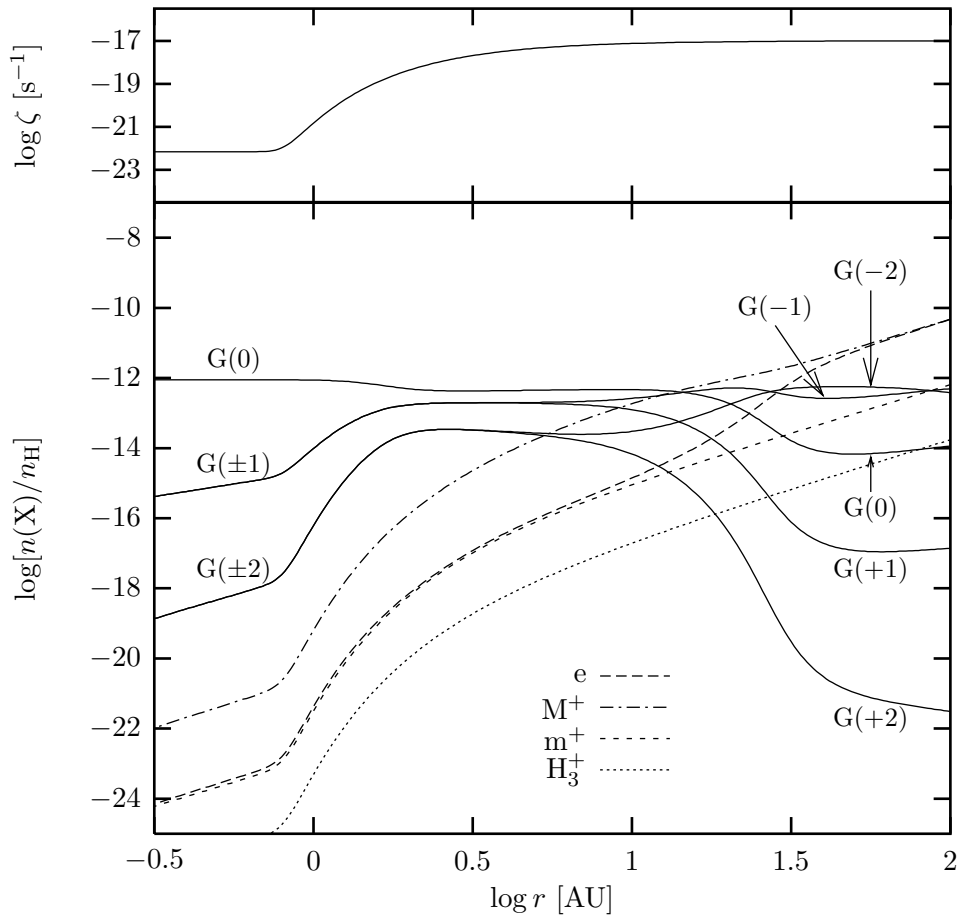
\begin{figure}
\begin{center}
\setlength{\unitlength}{0.1bp}
\begin{picture}(3600,1080)(0,0)
\put(100,615){%
\makebox(0,0)[b]{\shortstack{$\log \zeta ~[ {\rm s}^{-1} ]$}}%
}
\put(500,858){\makebox(0,0)[r]{$-17$}}
\put(500,696){\makebox(0,0)[r]{$-19$}}
\put(500,534){\makebox(0,0)[r]{$-21$}}
\put(500,372){\makebox(0,0)[r]{$-23$}}
\end{picture} \\
\vspace{-36pt}
\setlength{\unitlength}{0.1bp}
\begin{picture}(3600,2880)(0,0)
\put(2180,641){\makebox(0,0)[r]{H$_3^{+}$}}
\put(2180,759){\makebox(0,0)[r]{m$^{+}$}}
\put(2180,877){\makebox(0,0)[r]{M$^{+}$}}
\put(2180,995){\makebox(0,0)[r]{e}}
\put(3250,2516){\makebox(0,0){{\small G($-$2)}}}
\put(2890,2383){\makebox(0,0){{\small G($-$1)}}}
\put(3250,1623){\makebox(0,0){{\small G(0)}}}
\put(3250,830){\makebox(0,0){{\small G($+$2)}}}
\put(3250,1392){\makebox(0,0){{\small G($+$1)}}}
\put(850,1392){\makebox(0,0){{\small G($\pm$2)}}}
\put(850,1788){\makebox(0,0){{\small G($\pm$1)}}}
\put(850,2185){\makebox(0,0){{\small G(0)}}}
\put(2050,150){\makebox(0,0){$\log r$ [AU]}}
\put(100,1590){%
\makebox(0,0)[b]{\shortstack{$\log [ n({\rm X}) / n_{\rm H} ]$}}%
}
\put(3550,300){\makebox(0,0){$2$}}
\put(2950,300){\makebox(0,0){$1.5$}}
\put(2350,300){\makebox(0,0){$1$}}
\put(1750,300){\makebox(0,0){$0.5$}}
\put(1150,300){\makebox(0,0){$0$}}
\put(550,300){\makebox(0,0){$-0.5$}}
\put(500,2648){\makebox(0,0)[r]{$-8$}}
\put(500,2383){\makebox(0,0)[r]{$-10$}}
\put(500,2119){\makebox(0,0)[r]{$-12$}}
\put(500,1854){\makebox(0,0)[r]{$-14$}}
\put(500,1590){\makebox(0,0)[r]{$-16$}}
\put(500,1326){\makebox(0,0)[r]{$-18$}}
\put(500,1061){\makebox(0,0)[r]{$-20$}}
\put(500,797){\makebox(0,0)[r]{$-22$}}
\put(500,532){\makebox(0,0)[r]{$-24$}}
\end{picture} \\
\vspace{-12pt}
\caption
{Relative abundances $n({\rm X})/n_{\rm H}$ ({\it lower panel}) of some
representative particles on the midplane of a disk with $q = 3 / 2$,
$f_{\Sigma} = 1$, $p = 1 / 2$, $T_0 = 280$ K, $M_{\ast} = 1 ~M_{\odot}$,
and $f_{\rm g} = 1$ as functions of the distance $r$ from the star.  
We call this the fiducial model because it is expected to well represent
the primitive solar nebula in an early evolutionary stage.
We adopted the depletion factors of heavy elements from the gas phase
$\delta_1 = 0.2$ and $\delta_2 = 0.02$, the fraction of gas phase
oxygen in the form of O$_2$, H$_2$O, or OH $f_{{\rm O}_2} = 0.7$, and
the radius of dust grains $a = 0.1$ $\mu$m.
G($n$) represents grains with electric charge $ne$.
The upper panel shows the ionization rate $\zeta$ on the midplane as a
function of $r$.
\label{fig:rdep}}
\end{center}
\end{figure}

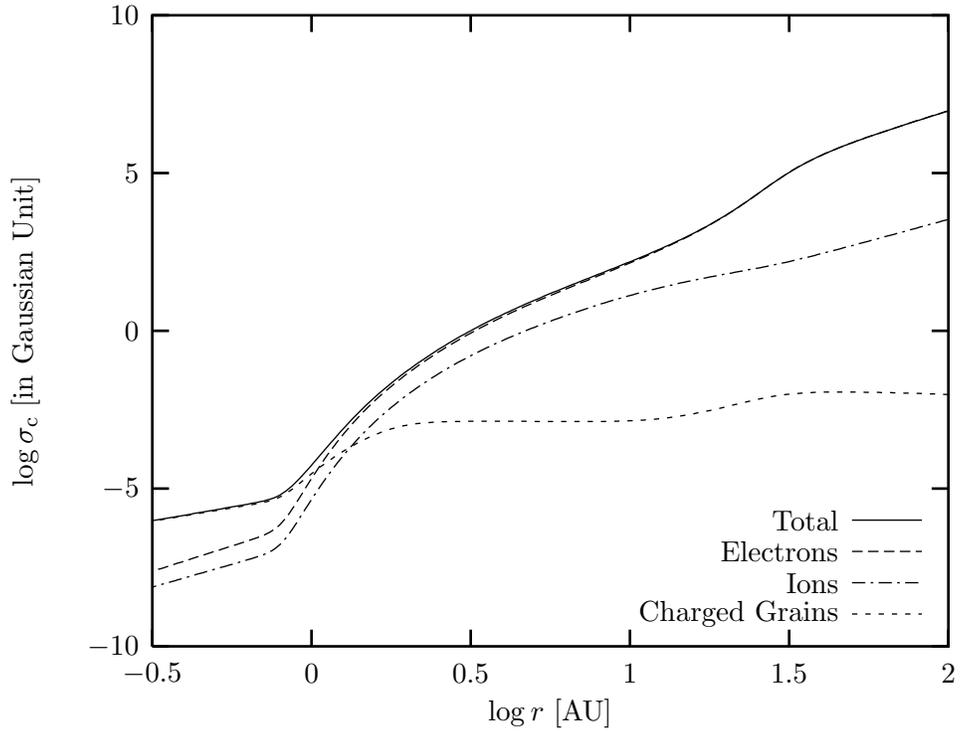
\begin{figure}
\begin{center}
\setlength{\unitlength}{0.1bp}
\begin{picture}(3600,2880)(0,0)
\put(3137,522){\makebox(0,0)[r]{Charged Grains}}
\put(3137,640){\makebox(0,0)[r]{Ions}}
\put(3137,758){\makebox(0,0)[r]{Electrons}}
\put(3137,876){\makebox(0,0)[r]{Total}}
\put(2050,150){\makebox(0,0){$\log r$ [AU]}}
\put(100,1590){%
\makebox(0,0)[b]{\shortstack{$\log \sigma_{\rm c}$ [in Gaussian Unit]}}%
}
\put(3550,300){\makebox(0,0){$2$}}
\put(2950,300){\makebox(0,0){$1.5$}}
\put(2350,300){\makebox(0,0){$1$}}
\put(1750,300){\makebox(0,0){$0.5$}}
\put(1150,300){\makebox(0,0){$0$}}
\put(550,300){\makebox(0,0){$-0.5$}}
\put(500,2780){\makebox(0,0)[r]{$10$}}
\put(500,2185){\makebox(0,0)[r]{$5$}}
\put(500,1590){\makebox(0,0)[r]{$0$}}
\put(500,995){\makebox(0,0)[r]{$-5$}}
\put(500,400){\makebox(0,0)[r]{$-10$}}
\end{picture} \\
\vspace{-12pt}
\caption
{The electrical conductivity $\sigma_{\rm c}$ on the midplane of the
fiducial model (same as in Fig.~{\protect{\ref{fig:rdep}}}) as a
function of the distance $r$ from the star.
Also shown are the contributions of electrons, ions, and charged
grains to the conductivity.
\label{fig:recond}}
\end{center}
\end{figure}

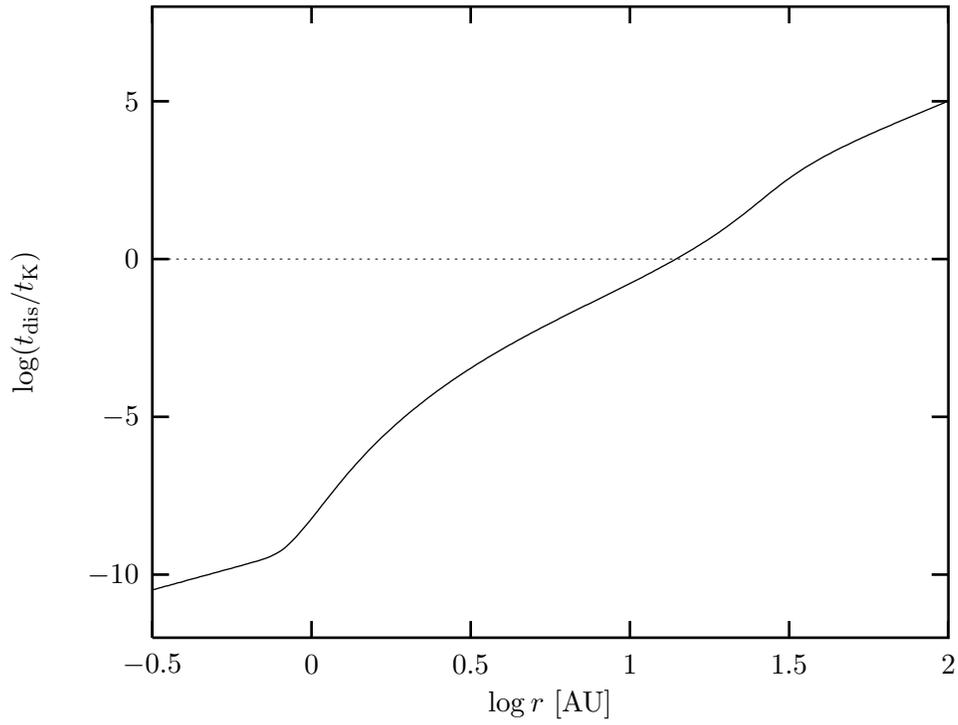
\begin{figure}
\begin{center}
\setlength{\unitlength}{0.1bp}
\begin{picture}(3600,2880)(0,0)
\put(2050,150){\makebox(0,0){$\log r$ [AU]}}
\put(100,1590){%
\makebox(0,0)[b]{\shortstack{$\log ( t_{\rm dis} / t_{\rm K} )$}}%
}
\put(3550,300){\makebox(0,0){$2$}}
\put(2950,300){\makebox(0,0){$1.5$}}
\put(2350,300){\makebox(0,0){$1$}}
\put(1750,300){\makebox(0,0){$0.5$}}
\put(1150,300){\makebox(0,0){$0$}}
\put(550,300){\makebox(0,0){$-0.5$}}
\put(500,2423){\makebox(0,0)[r]{$5$}}
\put(500,1828){\makebox(0,0)[r]{$0$}}
\put(500,1233){\makebox(0,0)[r]{$-5$}}
\put(500,638){\makebox(0,0)[r]{$-10$}}
\end{picture} \\
\vspace{-12pt}
\caption
{The ohmic dissipation time of magnetic field, $t_{\rm dis}$, on the
midplane relative to the Keplerian orbital period $t_{\rm K}$ as a
function of $r$ for the fiducial model.
The magnetic diffusivity $\eta$ given by
equation~({\protect{\ref{eqn:eta}}}) is evaluated with the abundances
of charged particles shown in Figure~{\protect{\ref{fig:rdep}}}.
\label{fig:rtd}}
\end{center}
\end{figure}

Figure~\ref{fig:rdep} shows the distribution of relative abundances $n
({\rm X}) / n_{\rm H}$ of some representative particles X on the
midplane for the fiducial model.  
The ionization rate at the midplane is also shown in the upper panel as
a function of the radius $r$.
Figure~\ref{fig:recond} shows the resulting electrical conductivity
$\sigma_{\rm c}$ given by equation~(\ref{eqn:cond}).
The abundances of charged particles and the electrical conductivity
decrease as the distance from the central star decreases.
When the time scale of magnetic field dissipation,
\begin{equation}
t_{\rm dis} \approx \frac{H^2}{\eta} \;,
\end{equation} 
is comparable to or smaller than the Keplerian orbital period $t_{\rm
K} = 2 \pi / \Omega$, magnetic fields are almost decoupled from the
gas.  
We call the distance from the central star at which $t_{\rm dis} =
t_{\rm K}$ holds the decoupling radius $r_{\rm dec}$.
Figure \ref{fig:rtd} depicts the ratio of the dissipation time $t_{\rm 
dis}$ to the dynamical time $t_{\rm K}$ as a function of $r$.
We find $t_{\rm dis} \lesssim t_{\rm K}$ at $r < r_{\rm dec} \approx
15$ AU, where the hydrogen number density $n_{\rm H}$ is higher than
$10^{12}~{\rm cm}^{-3}$.

In the region of $r \gtrsim r_{\rm dec}$, electrons and metal ions are
the dominant charged particles as seen from Figure~\ref{fig:rdep}. 
Electrons and ions recombine mainly by collision with dust grains, and
thus the relative abundances are inversely proportional to $n_{\rm
H}$.
The conductivity is mainly contributed by electrons, and is
proportional to $x({\rm e})$.

In the region of  $r \lesssim r_{\rm dec}$, on the other hand, the
number density of dust grains is higher than that of ions, and the
dust grains with charge $\pm e$ are the dominant charged particles.
Because the charge neutrality must be kept mainly by grains, or
$n({\rm G}(+1)) \approx n({\rm G}(-1))$, the number density of free
electrons $n({\rm e})$ becomes considerably lower than that of metal
ions $n({\rm M}^{+})$. 
The ratio  $n({\rm e})/n({\rm M}^{+})$ approaches the limiting value
$( m_{\rm e} / S_{\rm e} m_{\rm M} )^{1/2}$ (Umebayashi 1983; Nakano
1984) as $r$ decreases from 10 AU, where $m_{\rm e}$ and $m_{\rm M}$
are masses of electron and metallic ion, respectively. 
At $r \lesssim 7$ AU, the column density of the disk exceeds the
attenuation length of cosmic rays, $\chi_{\rm CR}$, and the abundance
of charged particles reduces considerably due to decrease of the
ionization rate.
In the region of $r \lesssim 2$ AU, where the terrestrial planets
orbit in the present solar system, the abundances of dust grains with
electric charge $\pm e$ decrease as $r$ decreases. 
This is due to recombination of adsorbed electrons and ions during
grain-grain collision.  
At $r \lesssim 1$ AU, where the number density of ions is more than
$10^4$ times smaller than that of charged grains, the contribution to
$\sigma_{\rm c}$ is dominated by charged grains as seen in
Figure~\ref{fig:recond}.

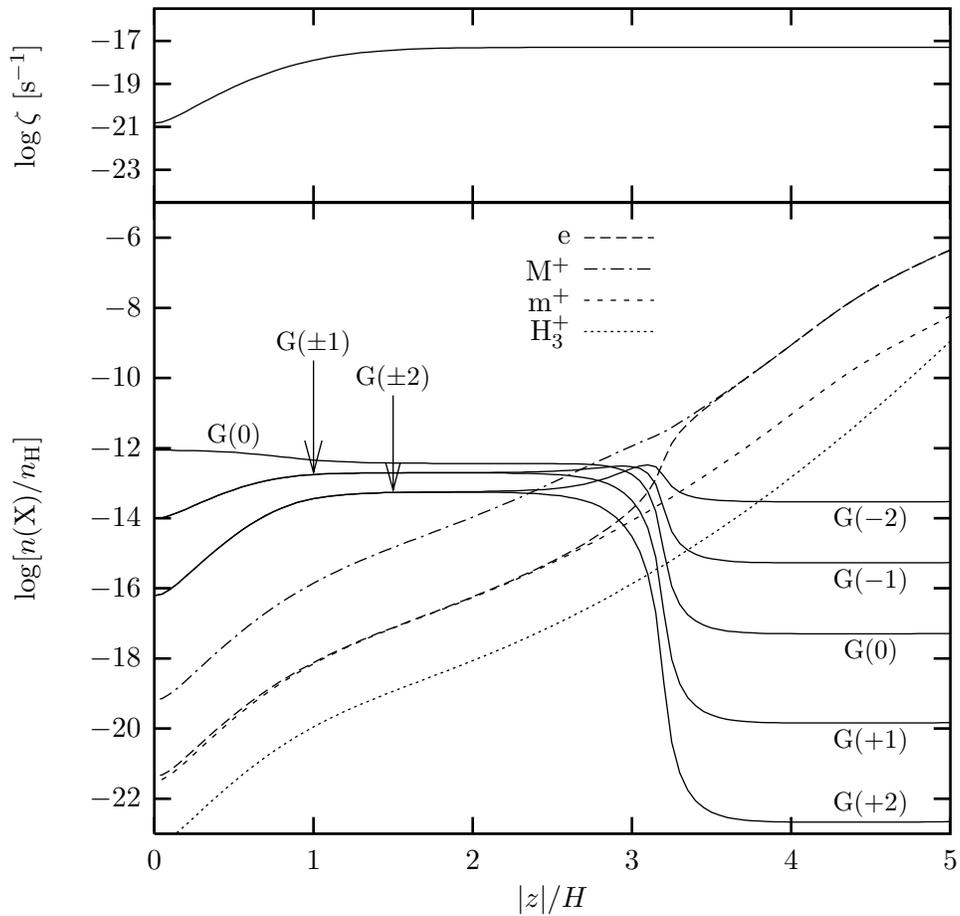
\begin{figure}
\begin{center}
\setlength{\unitlength}{0.1bp}
\begin{picture}(3600,1080)(0,0)
\put(100,615){%
\makebox(0,0)[b]{\shortstack{$\log \zeta ~[ {\rm s}^{-1} ]$}}%
}
\put(500,858){\makebox(0,0)[r]{$-17$}}
\put(500,696){\makebox(0,0)[r]{$-19$}}
\put(500,534){\makebox(0,0)[r]{$-21$}}
\put(500,372){\makebox(0,0)[r]{$-23$}}
\end{picture} \\
\vspace{-36pt}
\setlength{\unitlength}{0.1bp}
\begin{picture}(3600,2880)(0,0)
\put(2120,2294){\makebox(0,0)[r]{H$_3^{+}$}}
\put(2120,2412){\makebox(0,0)[r]{m$^{+}$}}
\put(2120,2530){\makebox(0,0)[r]{M$^{+}$}}
\put(2120,2648){\makebox(0,0)[r]{e}}
\put(3250,1583){\makebox(0,0){{\small G($-$2)}}}
\put(3250,1352){\makebox(0,0){{\small G($-$1)}}}
\put(3250,1088){\makebox(0,0){{\small G(0)}}}
\put(3250,512){\makebox(0,0){{\small G($+$2)}}}
\put(3250,750){\makebox(0,0){{\small G($+$1)}}}
\put(1450,2119){\makebox(0,0){{\small G($\pm$2)}}}
\put(1150,2251){\makebox(0,0){{\small G($\pm$1)}}}
\put(850,1901){\makebox(0,0){{\small G(0)}}}
\put(2050,150){\makebox(0,0){$|z| / H$}}
\put(100,1590){%
\makebox(0,0)[b]{\shortstack{$\log [ n({\rm X}) / n_{\rm H} ]$}}%
}
\put(3550,300){\makebox(0,0){$5$}}
\put(2950,300){\makebox(0,0){$4$}}
\put(2350,300){\makebox(0,0){$3$}}
\put(1750,300){\makebox(0,0){$2$}}
\put(1150,300){\makebox(0,0){$1$}}
\put(550,300){\makebox(0,0){$0$}}
\put(500,2648){\makebox(0,0)[r]{$-6$}}
\put(500,2383){\makebox(0,0)[r]{$-8$}}
\put(500,2119){\makebox(0,0)[r]{$-10$}}
\put(500,1854){\makebox(0,0)[r]{$-12$}}
\put(500,1590){\makebox(0,0)[r]{$-14$}}
\put(500,1326){\makebox(0,0)[r]{$-16$}}
\put(500,1061){\makebox(0,0)[r]{$-18$}}
\put(500,797){\makebox(0,0)[r]{$-20$}}
\put(500,532){\makebox(0,0)[r]{$-22$}}
\end{picture} \\
\vspace{-12pt}
\caption
{Relative abundances $n({\rm X})/n_{\rm H}$ of some representative
particles as functions of the height $|z|$ from the midplane at $r =
1$ AU for the same disk as in Figure~{\protect{\ref{fig:rdep}}}.
The upper panel shows the ionization rate $\zeta$ as a function of
$|z|$.
\label{fig:zdep}}
\end{center}
\end{figure}

We next investigate the abundances of charged particles and the
dissipation time of magnetic field off the midplane at some
representative regions.
Figure~\ref{fig:zdep} depicts the relative abundances of some
representative particles as functions of the height $|z|$ from the
midplane for the region $r = 1$ AU where the Earth may have formed.
As $|z|$ decreases, the abundances of charged particles decrease,
and as a result the electrical conductivity also decreases.
At $|z| \lesssim 3.0 ~H$, the number density of free electrons
deviates from that of metal ions, and dust grains with electric charge 
$\pm e$ are the dominant charged particles.  
At $|z| \lesssim 1.0 ~H$, the number density of charged grains
decreases as $|z|$ decreases because of the charge neutralization by
mutual collisions.
The conductivity is mainly contributed by ions and electrons except at
$|z| \lesssim 0.1 ~H$ where the number density of ions is lower than
$10^{-4}$ times that of charged grains. 

Figure~\ref{fig:zdep2} shows the relative abundances of some
representative particles for another typical region $r = 30$ AU where 
Neptune may have formed.
Because of the low density $n_{\rm H} \lesssim 10^{10} ~{\rm cm}^{-3}$,
electrons and metal ions are the dominant charged particles at any
height. 

Figure~\ref{fig:ztd} shows the dissipation time $t_{\rm dis}$ as a
function of $|z|$ at $r = 1$ and 30 AU.
In the region $r = 1$ AU $t_{\rm dis}$ is smaller than the dynamical
time $t_{\rm K}$ at $|z| \lesssim z_{\rm dec} \approx 3.2 ~H$ where the
number density $n_{H}$ exceeds $10^{11} ~{\rm cm}^{-3}$.
Notice that the column density of the well-coupled region, where
$t_{\rm dis} > t_{\rm K}$ holds, is smaller than $10^{-2} ~{\rm
g}~{\rm cm}^{-2}$, which is much less than the attenuation length of
cosmic rays $\chi_{\rm CR}$, and as a result $\zeta \approx \zeta_{\rm
CR}$ as shown in the upper panel of Figure~\ref{fig:zdep}, different
from Gammie (1996).
The low ionization fraction in the decoupled regions is caused merely
by the very high density, not by attenuation of cosmic rays.
At $r = 30$ AU we find $t_{\rm dis} \gg t_{\rm K}$ at any $z$, and
consequently the gas is strongly coupled with magnetic fields.

\subsubsection{Unstable Regions}

We shall find out the regions where the condition given by
equation~(\ref{eqn:rzcrit}) is satisfied using the spatial distribution
of magnetic diffusivity on the $r$-$z$ plane obtained above.
In such regions, the magnetorotational instability is operative (Paper
I), and thus we call them the unstable regions.

Since the stability condition also depends on the Alfv{\'e}n speed,
we have to specify the spatial distribution of magnetic field
strength.
For the sake of simplicity, we consider only the vertical component of 
magnetic fields, and assume that the plasma beta at the midplane,
$\beta_{\rm c} = 2 c_{\rm s}^2 / v_{\rm Ac}^2$, is constant in the disk,
where $v_{\rm Ac}$ is the Alfv{\'e}n speed at the midplane.
Then, the distribution of magnetic field strength for the disks with $q
= 3 / 2$, $M_{\ast} = 1 ~M_{\odot}$, $p = 1 / 2$, and $T_0 = 280$ K is
given by
\begin{equation}
B (r) = 1.9 \; f_{\Sigma}^{1/2} 
\left( \frac{r}{1 ~{\rm AU}} \right)^{-13/8}
\left( \frac{\beta_{\rm c}}{100} \right)^{-1/2} {\rm G} \;.
\end{equation}

\begin{figure}
\begin{center}
\setlength{\unitlength}{0.1bp}
\begin{picture}(3600,2880)(0,0)
\put(964,2347){\makebox(0,0)[r]{$H$}}
\put(2798,1302){\makebox(0,0){{\fbox{\bsf Unstable Region}}}}
\put(647,2564){\makebox(0,0)[l]{(a) $\beta_{\rm c} = 100$}}
\put(1624,1842){\makebox(0,0){$\lambda_{\rm ideal}/H = 1$}}
\put(920,1482){\makebox(0,0){$\lambda_{\rm res}/H = 1$}}
\put(2000,150){\makebox(0,0){$r$ [AU]}}
\put(100,1590){%
\makebox(0,0)[b]{\shortstack{$z$ [AU]}}%
}
\put(3268,300){\makebox(0,0){$30$}}
\put(2329,300){\makebox(0,0){$20$}}
\put(1389,300){\makebox(0,0){$10$}}
\put(450,300){\makebox(0,0){$0$}}
\put(400,2564){\makebox(0,0)[r]{$3$}}
\put(400,1842){\makebox(0,0)[r]{$2$}}
\put(400,1121){\makebox(0,0)[r]{$1$}}
\put(400,400){\makebox(0,0)[r]{$0$}}
\end{picture} \\
\setlength{\unitlength}{0.1bp}
\begin{picture}(3600,2880)(0,0)
\put(964,2347){\makebox(0,0)[r]{$H$}}
\put(2798,1842){\makebox(0,0){{\fbox{\bsf Unstable Region}}}}
\put(647,2564){\makebox(0,0)[l]{(b) $\beta_{\rm c} = 1000$}}
\put(1577,1842){\makebox(0,0)[r]{$\lambda_{\rm ideal}/H = 1$}}
\put(2460,761){\makebox(0,0)[r]{$\lambda_{\rm res}/H = 1$}}
\put(2000,150){\makebox(0,0){$r$ [AU]}}
\put(100,1590){%
\makebox(0,0)[b]{\shortstack{$z$ [AU]}}%
}
\put(3268,300){\makebox(0,0){$30$}}
\put(2329,300){\makebox(0,0){$20$}}
\put(1389,300){\makebox(0,0){$10$}}
\put(450,300){\makebox(0,0){$0$}}
\put(400,2564){\makebox(0,0)[r]{$3$}}
\put(400,1842){\makebox(0,0)[r]{$2$}}
\put(400,1121){\makebox(0,0)[r]{$1$}}
\put(400,400){\makebox(0,0)[r]{$0$}}
\end{picture} \\
\vspace{-12pt}
\caption
{The unstable regions ({\it striped}) for the cases of (a) the plasma beta at
the midplane $\beta_{\rm c} = 100$ and (b) $\beta_{\rm c} = 1000$ for
the fiducial model (same as in Fig.~{\protect{\ref{fig:rdep}}}).
The dotted curve denotes the scale height of the disk, $z = H (r)$.
In the region above the curve $\lambda_{\rm ideal} / H = 1$, the
instability does not effectively work because the wavelengths of the
unstable modes are larger than the scale height of the disk.
The region below the curve $\lambda_{\rm res} / H = 1$ is stable
because the magnetic dissipation is effective, and corresponds to
Gammie's (1996) ``dead zone''.
\label{fig:d100}}
\end{center}
\end{figure}
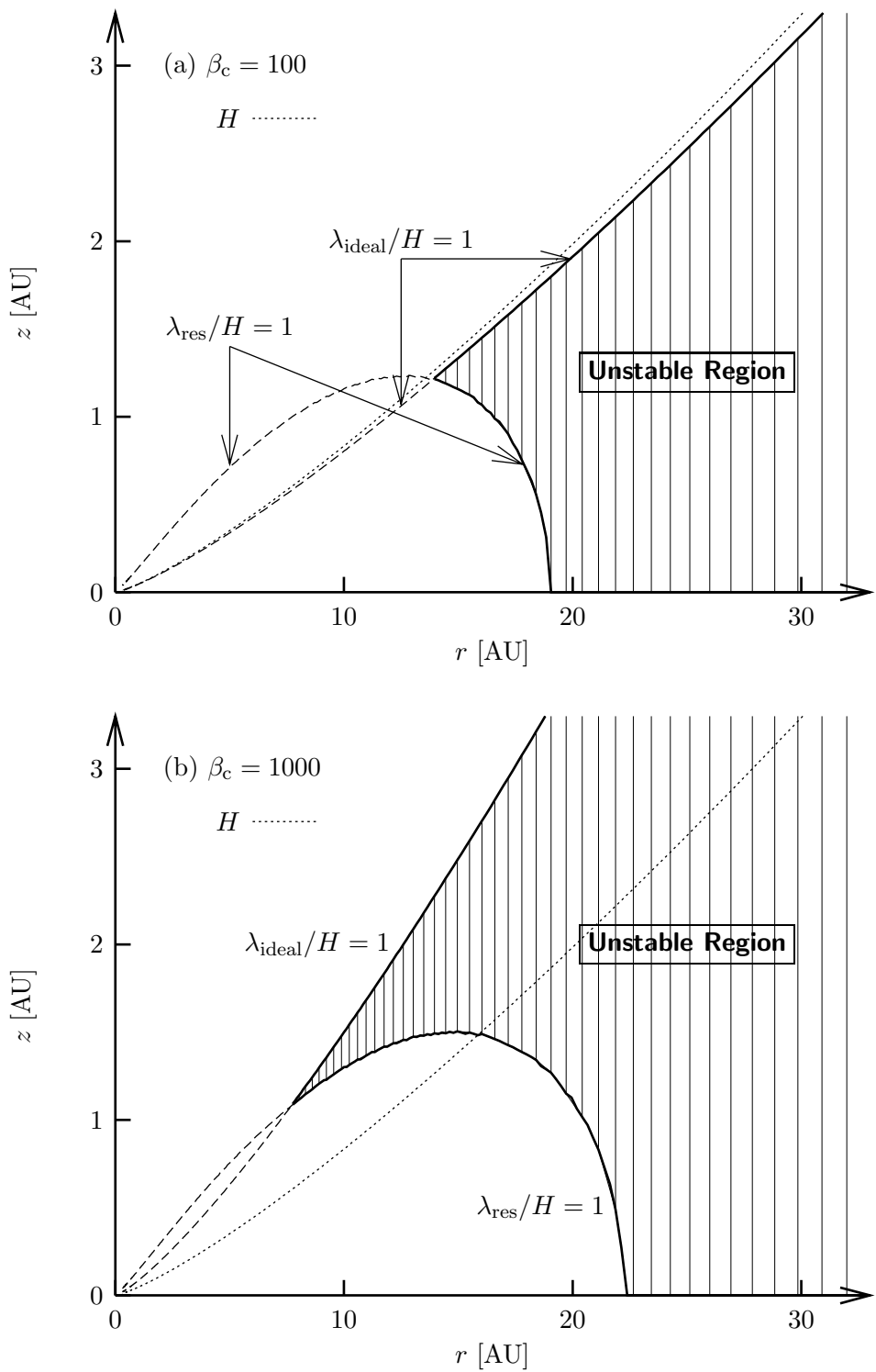

Figures~\ref{fig:d100}a and \ref{fig:d100}b show the unstable regions
for the cases of $\beta_{\rm c} = 100$ and 1000, respectively.  
In these figures the curve labeled $\lambda_{\rm ideal} / H = 1$
represents the locus of the points at which the characteristic unstable
wavelength in the ideal MHD limit, $\lambda_{\rm ideal}$, is equal to
the scale height of the disk,
\begin{equation}
\lambda_{\rm ideal} (r,z) \equiv 2 \pi \frac{v_{\rm A}
(r,z)}{\Omega(r)} = H (r) \;.
\label{eqn:lamii}
\end{equation}
The curve with $\lambda_{\rm res} / H = 1$ is the locus of the points at
which the characteristic unstable wavelength in the resistive limit,
$\lambda_{\rm res}$, is equal to the scale height,
\begin{equation}
\lambda_{\rm res} (r,z) \equiv 2 \pi \frac{\eta(r,z)}{v_{\rm A}
(r,z)} = H (r) \;.
\label{eqn:lamrr}
\end{equation}
The critical condition for decoupling of magnetic field, $t_{\rm dis}
= t_{\rm K}$, mentioned above is equivalent to
equation~(\ref{eqn:lamrr}) when $\beta_{\rm c} = 1$.

For a given radius, $\lambda_{\rm ideal}$ increases exponentially as
the height $|z|$ increases because the Alfv{\'e}n speed becomes greater, 
while $\lambda_{\rm res}$ decreases because the magnetic diffusivity
decreases as $|z|$ increases.  
Therefore, the instability condition $\max ( \lambda_{\rm ideal} ,
\lambda_{\rm res} ) \lesssim H$ [eq.~(\ref{eqn:rzcrit})] is satisfied
only in the striped area between the two critical curves $\lambda_{\rm
ideal} / H = 1$ and $\lambda_{\rm res} / H = 1$ in
Figure~\ref{fig:d100}.

The condition $\lambda_{\rm ideal} / H = 1$ is easily reduced to
\begin{equation}
| z | = 
\displaystyle \left( \ln \frac{\beta_c}{4 \pi^2} \right)^{1/2} H \;,
\end{equation}
which gives $| z | = 0.96 ~H$ and $1.8 ~H$ for $\beta_{\rm c} =$ 100
and 1000, respectively. 
The magnetorotational instability is suppressed due to magnetic
dissipation in the region below the critical curve $\lambda_{\rm res}
/ H = 1$, which corresponds to Gammie's (1996) ``dead zone''.
The dissipation process is less important for stronger magnetic field.
Thus the stable region shrinks as the plasma beta decreases.
At the points where $\lambda_{\rm res} / H = 1$ holds, the hydrogen
number density $n_{\rm H}$ takes 0.8 -- $1.7 \times 10^{11}$ and 0.5
-- $1.2 \times 10^{11}$ cm$^{-3}$ for the cases of $\beta_{\rm c} =
100$ and 1000, respectively.
At temperatures below several hundred Kelvin where thermal ionization
is negligible, the magnetic diffusivity is determined mainly by the
density and the ionization rate.
Because the column density above the dead zone is far below the
attenuation length of cosmic rays, the boundary of the dead zone is
determined mostly by the local density in this case.

The electrical conductivity is mainly contributed by electrons at the
boundary of the dead zone.
The relative abundance of electrons $x ({\rm e})$ at the boundary is
$1.1$ -- $2.7 \times 10^{-14}$ and $2.8$ -- $7.1 \times 10^{-14}$ for
the cases of $\beta_{\rm c} = 100$ and 1000, respectively.
The magnetic diffusivity is inversely proportional to $x ({\rm e})$ in 
such a situation, and then we approximately have $x ({\rm e}) \propto
v_{\rm A}^{-1}$ on the curve $\lambda_{\rm res} / H = 1$ at a given
radius.

Let the critical radius $r_{\rm crit}$ be the outer radius of the dead 
zone on the midplane, which is about 19 and 22 AU for the cases of
$\beta_{\rm c} = 100$ and 1000, respectively.  
The region $r \gtrsim r_{\rm crit}$ is magnetorotationally unstable
except at tenuous outer layers.  
In this region, because the magnetic field is amplified due to growth
of fluctuations, the angular momentum can be transported effectively.
As seen from Figure~\ref{fig:d100}b, the unstable region exists even
at $r < r_{\rm crit}$ above the dead zone.
However, the column density of this unstable layer is so small that
mass accretion cannot be as active as in the region $r > r_{\rm
crit}$.
Therefore, the accreting matter from the outer region must accumulate
near the outer boundary of the dead zone, unless some other mechanisms
of angular momentum transport are effective.

In the following subsections, we shall depict the unstable regions on
the $r$-$z$ plane for various disk models.  
The criterion given by equation~(\ref{eqn:lamrr}) is more important
than the criterion~(\ref{eqn:lamii}) because the critical radius on
the midplane $r_{\rm crit}$ is determined by
equation~(\ref{eqn:lamrr}).
Therefore, in the following we focus on the dead zone given by the
criterion $\lambda_{\rm res} / H = 1$.
Because the results are not very sensitive to the plasma beta as seen
from Figure~\ref{fig:d100}, we show only the results for the cases of
$\beta_{\rm c} = 100$ and 1000 in the following.

\subsection{Disks with Various Surface Densities}

\subsubsection{Dependence on the Disk Mass}

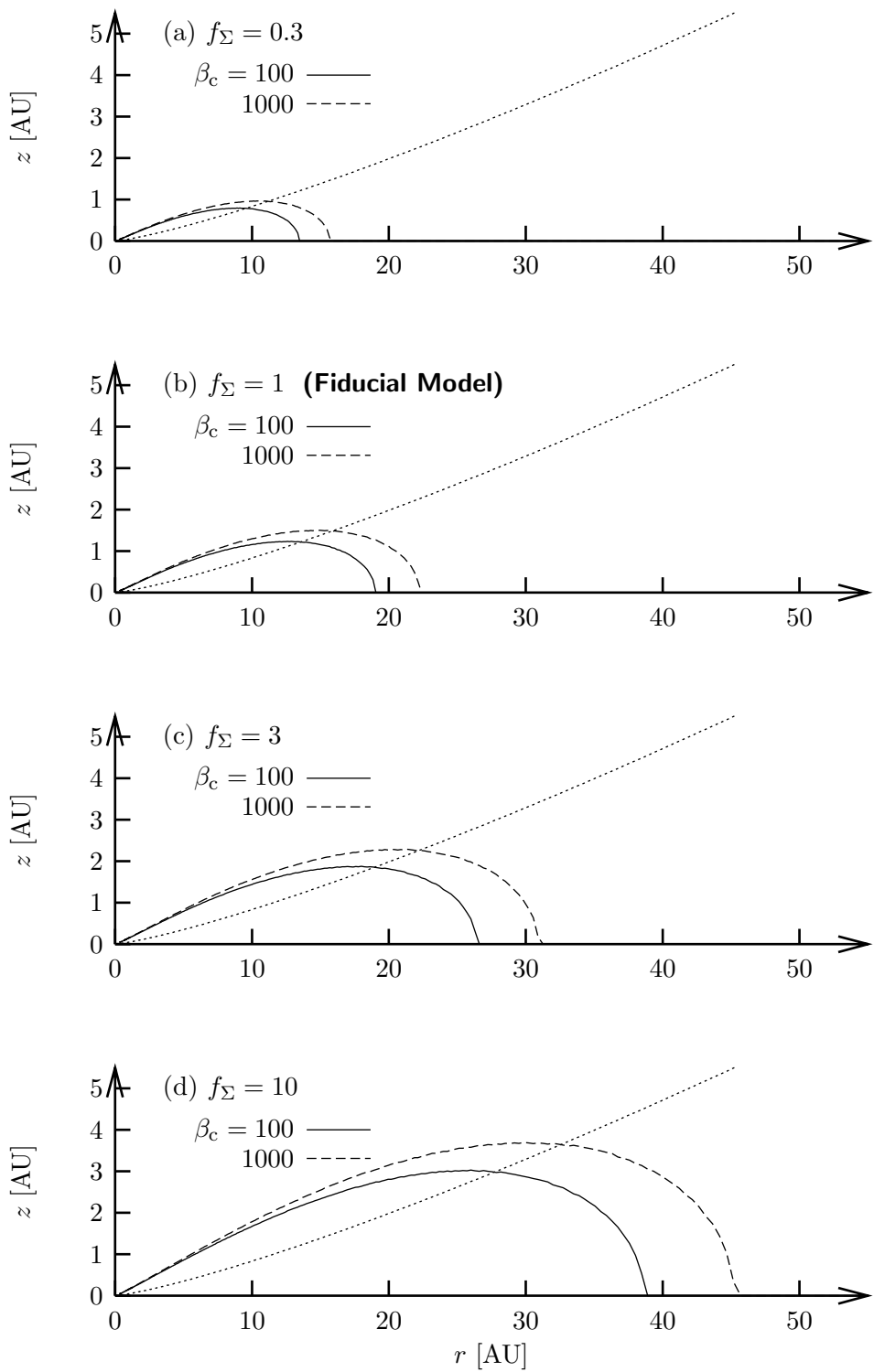
\begin{figure}
\begin{center}
\setlength{\unitlength}{0.1bp}
\begin{picture}(3600,1440)(0,0)
\put(1189,966){\makebox(0,0)[r]{1000}}
\put(1189,1084){\makebox(0,0)[r]{$\beta_{\rm c} = 100$}}
\put(647,1255){\makebox(0,0)[l]{(a) $f_{\Sigma} = 0.3$}}
\put(2000,150){\makebox(0,0){ }}
\put(100,870){%
\makebox(0,0)[b]{\shortstack{$z$ [AU]}}%
}
\put(3268,300){\makebox(0,0){$50$}}
\put(2705,300){\makebox(0,0){$40$}}
\put(2141,300){\makebox(0,0){$30$}}
\put(1577,300){\makebox(0,0){$20$}}
\put(1014,300){\makebox(0,0){$10$}}
\put(450,300){\makebox(0,0){$0$}}
\put(400,1255){\makebox(0,0)[r]{$5$}}
\put(400,1084){\makebox(0,0)[r]{$4$}}
\put(400,913){\makebox(0,0)[r]{$3$}}
\put(400,742){\makebox(0,0)[r]{$2$}}
\put(400,571){\makebox(0,0)[r]{$1$}}
\put(400,400){\makebox(0,0)[r]{$0$}}
\end{picture} \\
\setlength{\unitlength}{0.1bp}
\begin{picture}(3600,1440)(0,0)
\put(1189,966){\makebox(0,0)[r]{1000}}
\put(1189,1084){\makebox(0,0)[r]{$\beta_{\rm c} = 100$}}
\put(647,1255){\makebox(0,0)[l]{(b) $f_{\Sigma} = 1$ ~{\bsf (Fiducial Model)}}}
\put(2000,150){\makebox(0,0){ }}
\put(100,870){%
\makebox(0,0)[b]{\shortstack{$z$ [AU]}}%
}
\put(3268,300){\makebox(0,0){$50$}}
\put(2705,300){\makebox(0,0){$40$}}
\put(2141,300){\makebox(0,0){$30$}}
\put(1577,300){\makebox(0,0){$20$}}
\put(1014,300){\makebox(0,0){$10$}}
\put(450,300){\makebox(0,0){$0$}}
\put(400,1255){\makebox(0,0)[r]{$5$}}
\put(400,1084){\makebox(0,0)[r]{$4$}}
\put(400,913){\makebox(0,0)[r]{$3$}}
\put(400,742){\makebox(0,0)[r]{$2$}}
\put(400,571){\makebox(0,0)[r]{$1$}}
\put(400,400){\makebox(0,0)[r]{$0$}}
\end{picture} \\
\setlength{\unitlength}{0.1bp}
\begin{picture}(3600,1440)(0,0)
\put(1189,966){\makebox(0,0)[r]{1000}}
\put(1189,1084){\makebox(0,0)[r]{$\beta_{\rm c} = 100$}}
\put(647,1255){\makebox(0,0)[l]{(c) $f_{\Sigma} = 3$}}
\put(2000,150){\makebox(0,0){ }}
\put(100,870){%
\makebox(0,0)[b]{\shortstack{$z$ [AU]}}%
}
\put(3268,300){\makebox(0,0){$50$}}
\put(2705,300){\makebox(0,0){$40$}}
\put(2141,300){\makebox(0,0){$30$}}
\put(1577,300){\makebox(0,0){$20$}}
\put(1014,300){\makebox(0,0){$10$}}
\put(450,300){\makebox(0,0){$0$}}
\put(400,1255){\makebox(0,0)[r]{$5$}}
\put(400,1084){\makebox(0,0)[r]{$4$}}
\put(400,913){\makebox(0,0)[r]{$3$}}
\put(400,742){\makebox(0,0)[r]{$2$}}
\put(400,571){\makebox(0,0)[r]{$1$}}
\put(400,400){\makebox(0,0)[r]{$0$}}
\end{picture} \\
\setlength{\unitlength}{0.1bp}
\begin{picture}(3600,1440)(0,0)
\put(1189,966){\makebox(0,0)[r]{1000}}
\put(1189,1084){\makebox(0,0)[r]{$\beta_{\rm c} = 100$}}
\put(647,1255){\makebox(0,0)[l]{(d) $f_{\Sigma} = 10$}}
\put(2000,150){\makebox(0,0){$r$ [AU]}}
\put(100,870){%
\makebox(0,0)[b]{\shortstack{$z$ [AU]}}%
}
\put(3268,300){\makebox(0,0){$50$}}
\put(2705,300){\makebox(0,0){$40$}}
\put(2141,300){\makebox(0,0){$30$}}
\put(1577,300){\makebox(0,0){$20$}}
\put(1014,300){\makebox(0,0){$10$}}
\put(450,300){\makebox(0,0){$0$}}
\put(400,1255){\makebox(0,0)[r]{$5$}}
\put(400,1084){\makebox(0,0)[r]{$4$}}
\put(400,913){\makebox(0,0)[r]{$3$}}
\put(400,742){\makebox(0,0)[r]{$2$}}
\put(400,571){\makebox(0,0)[r]{$1$}}
\put(400,400){\makebox(0,0)[r]{$0$}}
\end{picture} \\
\vspace{-12pt}
\caption
{Unstable regions in the disks with various surface densities;
(a) $f_{\Sigma} = 0.3$, (b) $f_{\Sigma} = 1$ (the minimum-mass solar
nebula), (c) $f_{\Sigma} = 3$, and (d) $f_{\Sigma} = 10$.
The other parameters are the same as those of the fiducial model.
The solid and dashed curves represent the loci of $\lambda_{\rm res} /
H = 1$ for the cases of the field strength $\beta_{\rm c} = 100$ and
1000, respectively, inside of which is magnetorotationally stable.
The dotted curve represents the scale height of the disk $z = H (r)$.
\label{fig:fs}}
\end{center}
\end{figure}

We investigate the disks with different surface densities rotating
around a star of $M_{\ast} = 1 ~M_{\odot}$.
We consider the following cases of the surface density parameter; (a)
$f_{\Sigma} = 0.3$, (b) $f_{\Sigma} = 1$ (the minimum-mass solar
nebula), (c) $f_{\Sigma} = 3$, and (d) $f_{\Sigma} = 10$.
Figure~\ref{fig:fs} shows the results for the cases with the
depletion factor of dust grains $f_{\rm g} =1$ corresponding to early
evolutionary stages of the disks. 
The other parameters are the same as in the fiducial model.
The solid and dashed curves in each panel represent the loci of
$\lambda_{\rm res} / H = 1$ for the cases of $\beta_{\rm c} =$ 100 and
1000, respectively.
Inside these curves are the dead zones where the ohmic dissipation
suppresses the magnetorotational instability.
The dotted curve shows the scale height of the disk $z = H(r)$. 

Because the dead zone has a height close to the scale height $H$ for
most part of $r < r_{\rm crit}$, most of the matter at $r < r_{\rm
crit}$ is in the region where the instability is suppressed.
Because the ohmic dissipation is not so efficient as to suppress the
magnetorotational instability outside the dead zones in
Figure~\ref{fig:fs}, magnetic fields are amplified to some extent by the
instability and angular momentum transport due to magnetic stress is
expected though mass accretion through the layers above the dead zones
may not be efficient because of their low column densities.

We find that a higher mass disk is more stable for the
magnetorotational instability; as the surface density parameter
$f_{\Sigma}$ increases, the dead zone expands and the critical radius
$r_{\rm crit}$ increases.  
This is because the recombination processes are more effective in
denser regions, and thus the ionization fraction is lower in higher
mass disks.
In the case of $\beta_{\rm c} = 100$ (1000), the critical radius on
the midplane, $r_{\rm crit}$, is 13 (16), 19 (22), 26 (31), and 39 (45)
AU for $f_{\Sigma} = 0.3$, 1, 3, and 10, respectively.  
At the points where $\lambda_{\rm res} / H = 1$ holds, the hydrogen
number density takes a value between a few $\times~10^{10}$ and a few
$\times~10^{11}$ cm$^{-3}$ almost independent of $f_{\Sigma}$.
Therefore, the boundary of the dead zone is determined mainly by the
local density $n_{\rm H}$; cosmic rays are hardly attenuated at these
points as in the fiducial model.

\subsubsection{Dependence on the Power Index of $\Sigma$}

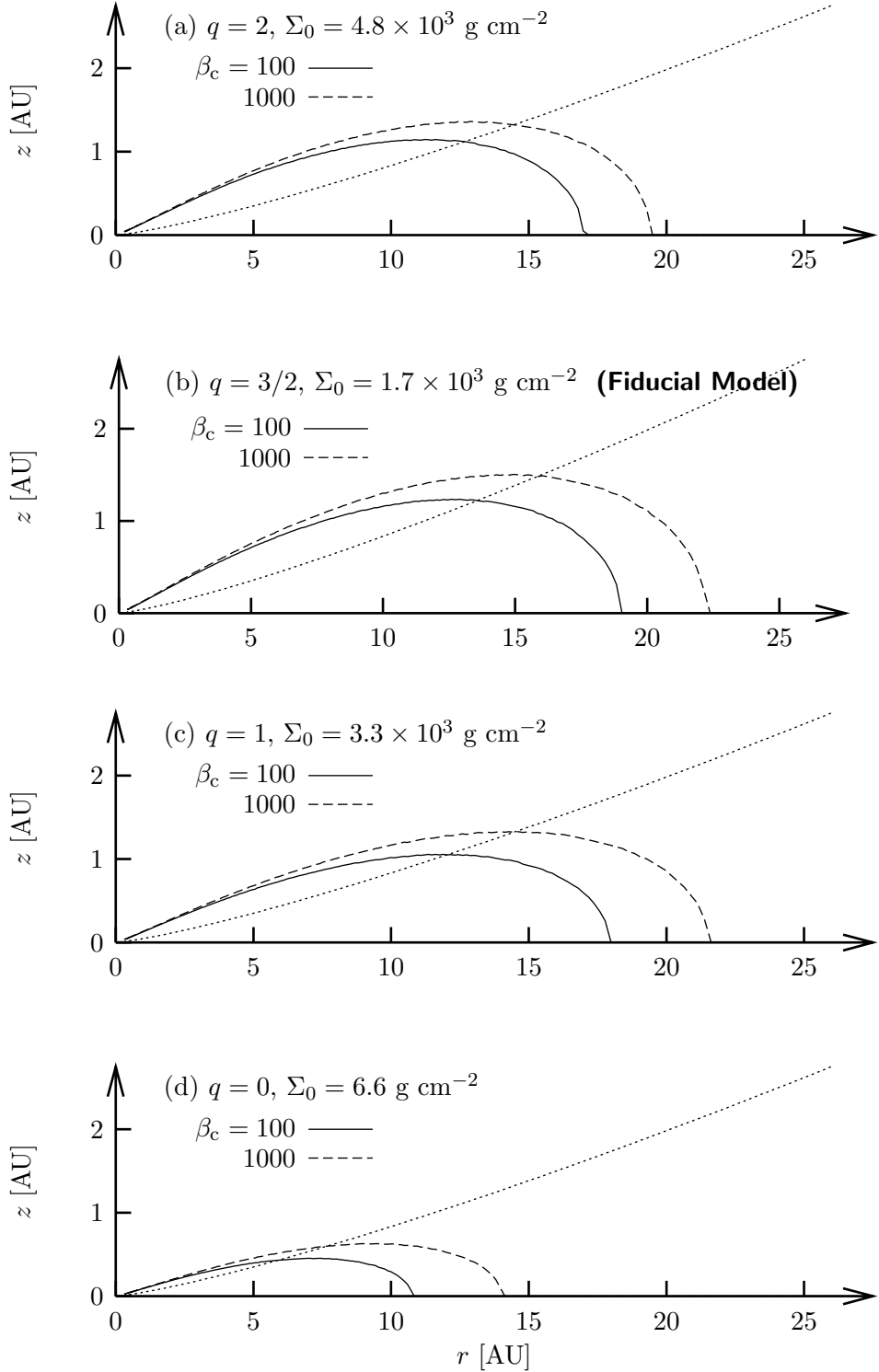
\begin{figure}
\begin{center}
\setlength{\unitlength}{0.1bp}
\begin{picture}(3600,1440)(0,0)
\put(1189,966){\makebox(0,0)[r]{1000}}
\put(1189,1084){\makebox(0,0)[r]{$\beta_{\rm c} = 100$}}
\put(647,1255){\makebox(0,0)[l]{(a) $q = 2$, $\Sigma_0 = 4.8 \times 10^3$ g~cm$^{-2}$}}
\put(2000,150){\makebox(0,0){ }}
\put(100,870){%
\makebox(0,0)[b]{\shortstack{$z$ [AU]}}%
}
\put(3268,300){\makebox(0,0){$25$}}
\put(2705,300){\makebox(0,0){$20$}}
\put(2141,300){\makebox(0,0){$15$}}
\put(1577,300){\makebox(0,0){$10$}}
\put(1014,300){\makebox(0,0){$5$}}
\put(450,300){\makebox(0,0){$0$}}
\put(400,1084){\makebox(0,0)[r]{$2$}}
\put(400,742){\makebox(0,0)[r]{$1$}}
\put(400,400){\makebox(0,0)[r]{$0$}}
\end{picture} \\
\setlength{\unitlength}{0.1bp}
\begin{picture}(3600,1440)(0,0)
\put(1170,938){\makebox(0,0)[r]{1000}}
\put(1170,1056){\makebox(0,0)[r]{$\beta_{\rm c} = 100$}}
\put(652,1245){\makebox(0,0)[l]{(b) $q = 3 / 2$, $\Sigma_0 = 1.7 \times 10^3$ g~cm$^{-2}$ ~{\bsf (Fiducial Model)}}}
\put(1950,100){\makebox(0,0){ }}
\put(100,820){%
\makebox(0,0)[b]{\shortstack{$z$ [AU]}}%
}
\put(3167,200){\makebox(0,0){$25$}}
\put(2626,200){\makebox(0,0){$20$}}
\put(2085,200){\makebox(0,0){$15$}}
\put(1544,200){\makebox(0,0){$10$}}
\put(1004,200){\makebox(0,0){$5$}}
\put(463,200){\makebox(0,0){$0$}}
\put(413,1056){\makebox(0,0)[r]{$2$}}
\put(413,678){\makebox(0,0)[r]{$1$}}
\put(413,300){\makebox(0,0)[r]{$0$}}
\end{picture} \\
\setlength{\unitlength}{0.1bp}
\begin{picture}(3600,1440)(0,0)
\put(1189,966){\makebox(0,0)[r]{1000}}
\put(1189,1084){\makebox(0,0)[r]{$\beta_{\rm c} = 100$}}
\put(647,1255){\makebox(0,0)[l]{(c) $q = 1$, $\Sigma_0 = 3.3 \times 10^3$ g~cm$^{-2}$}}
\put(2000,150){\makebox(0,0){ }}
\put(100,870){%
\makebox(0,0)[b]{\shortstack{$z$ [AU]}}%
}
\put(3268,300){\makebox(0,0){$25$}}
\put(2705,300){\makebox(0,0){$20$}}
\put(2141,300){\makebox(0,0){$15$}}
\put(1577,300){\makebox(0,0){$10$}}
\put(1014,300){\makebox(0,0){$5$}}
\put(450,300){\makebox(0,0){$0$}}
\put(400,1084){\makebox(0,0)[r]{$2$}}
\put(400,742){\makebox(0,0)[r]{$1$}}
\put(400,400){\makebox(0,0)[r]{$0$}}
\end{picture} \\
\setlength{\unitlength}{0.1bp}
\begin{picture}(3600,1440)(0,0)
\put(1189,966){\makebox(0,0)[r]{1000}}
\put(1189,1084){\makebox(0,0)[r]{$\beta_{\rm c} = 100$}}
\put(647,1255){\makebox(0,0)[l]{(d) $q = 0$, $\Sigma_0 = 6.6$ g~cm$^{-2}$}}
\put(2000,150){\makebox(0,0){$r$ [AU]}}
\put(100,870){%
\makebox(0,0)[b]{\shortstack{$z$ [AU]}}%
}
\put(3268,300){\makebox(0,0){$25$}}
\put(2705,300){\makebox(0,0){$20$}}
\put(2141,300){\makebox(0,0){$15$}}
\put(1577,300){\makebox(0,0){$10$}}
\put(1014,300){\makebox(0,0){$5$}}
\put(450,300){\makebox(0,0){$0$}}
\put(400,1084){\makebox(0,0)[r]{$2$}}
\put(400,742){\makebox(0,0)[r]{$1$}}
\put(400,400){\makebox(0,0)[r]{$0$}}
\end{picture} \\
\vspace{-12pt}
\caption
{Unstable regions in the disks with various power indices for the
surface density; 
(a) $q = 2$, $\Sigma_0 = 4.8 \times 10^3$ g~cm$^{-2}$,
(b) $q = 3 / 2$, $\Sigma_0 = 1.7 \times 10^3$ g~cm$^{-2}$, 
(c) $q = 1$, $\Sigma_0 = 3.3 \times 10^2$ g~cm$^{-2}$, and 
(d) $q = 0$, $\Sigma_0 = 6.6$ g~cm$^{-2}$. 
The other parameters are the same as those of the fiducial model.
All these models have the same disk mass 0.024 $M_{\odot}$ between 0.1
and 100 AU.
The solid and dashed curves represent the loci of $\lambda_{\rm res} /
H = 1$ for the cases of the field strength $\beta_{\rm c} = 100$ and
1000, respectively, inside of which is magnetorotationally stable.
The dotted curve represents the scale height of the disk $z = H (r)$.
\label{fig:q}}
\end{center}
\end{figure}

So far we have considered only the disks with the power index of
$\Sigma$, $q = 3 / 2$.
There are no definite observations that constrain surface density
profiles of protoplanetary disks.
Some of the extrasolar planet candidates discovered by radial velocity
oscillation orbit within just 0.3 AU, suggesting that they have migrated 
inward after their birth (e.g., Marcy, Cochran, and Mayor 2000).
Thus the mass distribution of the present-day solar system might be
different from the primordial distribution.
Here we examine some other models with different power index; (a) $q =
2$, (b) $q = 3 / 2$ (fiducial model), (c) $q = 1$, and (d) $q = 0$.
We assume that all these models have equal mass 0.024 $M_{\odot}$
between 0.1 and 100 AU.
Thus the surface density at 1 AU is $4.8 \times 10^{3}$, $1.7 \times
10^{3}$, $3.3 \times 10^{2}$, and $6.6$ g~cm$^{-2}$ for $q = 2$, 3/2, 1,
and 0, respectively.
The results for a disk of $f_{\rm g} = 1$ around a star of $M_{\ast} = 1
~M_{\odot}$ are shown in Figure~\ref{fig:q}.
The other parameters are the same as in the fiducial model.
The results are quite similar among these models though the critical
radius $r_{\rm crit}$ takes somewhat different values.
The critical radius on the midplane is mainly determined by the density
which is about $1.6 \times 10^{11}$ and $1.1 \times 10^{11}$ cm$^{-3}$
for $\beta_{\rm c} = 100$ and 1000, respectively.

\subsection{Disks in Some Evolutionary Stages}

\subsubsection{Depletion of Dust Grains}

The ionization fraction and the species of dominant charged particles
are sensitive to the depletion factor of dust grains, $f_{\rm g}$,
the abundance of dust grains relative to that in molecular clouds.
As the dust grains in a protoplanetary disk sink gradually toward the
midplane, $f_{\rm g}$ decreases and the recombination rate on grain
surfaces reduces except in a very thin layer around the midplane.
As a result, the ionization fraction increases (Umebayashi \& Nakano
1988) and the dead zone shrinks in the late evolutionary stages.

\begin{figure}
\begin{center}
\setlength{\unitlength}{0.1bp}
\begin{picture}(3600,1080)(0,0)
\put(100,615){%
\makebox(0,0)[b]{\shortstack{$\log \zeta ~[ {\rm s}^{-1} ]$}}%
}
\put(500,858){\makebox(0,0)[r]{$-17$}}
\put(500,696){\makebox(0,0)[r]{$-19$}}
\put(500,534){\makebox(0,0)[r]{$-21$}}
\put(500,372){\makebox(0,0)[r]{$-23$}}
\end{picture} \\
\vspace{-36pt}
\setlength{\unitlength}{0.1bp}
\begin{picture}(3600,2880)(0,0)
\put(980,2294){\makebox(0,0)[r]{H$_3^{+}$}}
\put(980,2412){\makebox(0,0)[r]{m$^{+}$}}
\put(980,2530){\makebox(0,0)[r]{M$^{+}$}}
\put(980,2648){\makebox(0,0)[r]{e}}
\put(1450,532){\makebox(0,0){G($+$1)}}
\put(3250,549){\makebox(0,0){G(0)}}
\put(3250,816){\makebox(0,0){G($-$1)}}
\put(3250,1048){\makebox(0,0){G($-$2)}}
\put(2050,2119){\makebox(0,0)[l]{e, M$^{+}$}}
\put(2050,150){\makebox(0,0){$|z| / H$}}
\put(100,1590){%
\makebox(0,0)[b]{\shortstack{$\log [ n({\rm X}) / n_{\rm H} ]$}}%
}
\put(3550,300){\makebox(0,0){$5$}}
\put(2950,300){\makebox(0,0){$4$}}
\put(2350,300){\makebox(0,0){$3$}}
\put(1750,300){\makebox(0,0){$2$}}
\put(1150,300){\makebox(0,0){$1$}}
\put(550,300){\makebox(0,0){$0$}}
\put(500,2648){\makebox(0,0)[r]{$-6$}}
\put(500,2383){\makebox(0,0)[r]{$-8$}}
\put(500,2119){\makebox(0,0)[r]{$-10$}}
\put(500,1854){\makebox(0,0)[r]{$-12$}}
\put(500,1590){\makebox(0,0)[r]{$-14$}}
\put(500,1326){\makebox(0,0)[r]{$-16$}}
\put(500,1061){\makebox(0,0)[r]{$-18$}}
\put(500,797){\makebox(0,0)[r]{$-20$}}
\put(500,532){\makebox(0,0)[r]{$-22$}}
\end{picture} \\
\vspace{-12pt}
\caption
{Relative abundances $n({\rm X})/n_{\rm H}$ of some representative
particles for the case of $f_{\rm g} = 10^{-4}$ as functions of the
height $|z|$ from the midplane at $r = 1$ AU.  
The other parameters are the same as those of the fiducial model.
The upper panel shows the ionization rate $\zeta$.
\label{fig:zdep-fg}}
\end{center}
\end{figure}
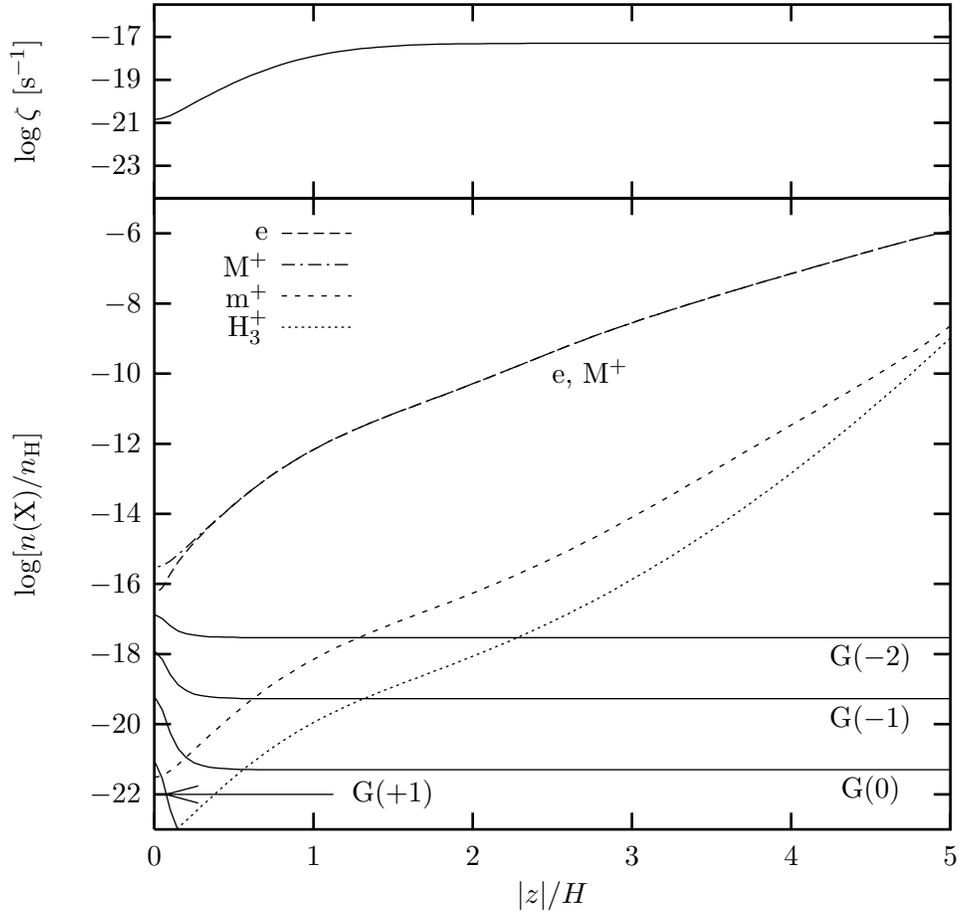

Figure~\ref{fig:zdep-fg} shows the relative abundances of some
representative particles as functions of the height from the midplane
at $r = 1$ AU for the case of $f_{\rm g} = 10^{-4}$. 
Electrons and metal ions are the dominant charged particles at any
height $z$ due to the low abundance of dust grains.
Magnetic fields are well coupled with the gas in the larger part of
the disk compared with the case of $f_{\rm g} = 1$.
The electrical conductivity $\sigma_{\rm c}$ is mainly contributed by
electrons, and is proportional to $x ({\rm e})$ at any $z$.
At $|z| \lesssim 1.0 ~H$, the ionization by cosmic rays declines
(upper panel of Fig.~\ref{fig:zdep-fg}) because the depth $\chi
(r,z)$ exceeds the attenuation length of cosmic rays, $\chi_{\rm CR}$.
As seen from the dot-dashed curve in Figure~\ref{fig:ztd}, the ratio
$t_{\rm dis} / t_{\rm K}$ becomes less than unity at $|z| \lesssim
z_{\rm dec} \approx 0.8 ~H$, so that the gas decouples from magnetic
fields in this region.
This decoupled layer is significantly thinner than for the case of
$f_{\rm g} = 1$.

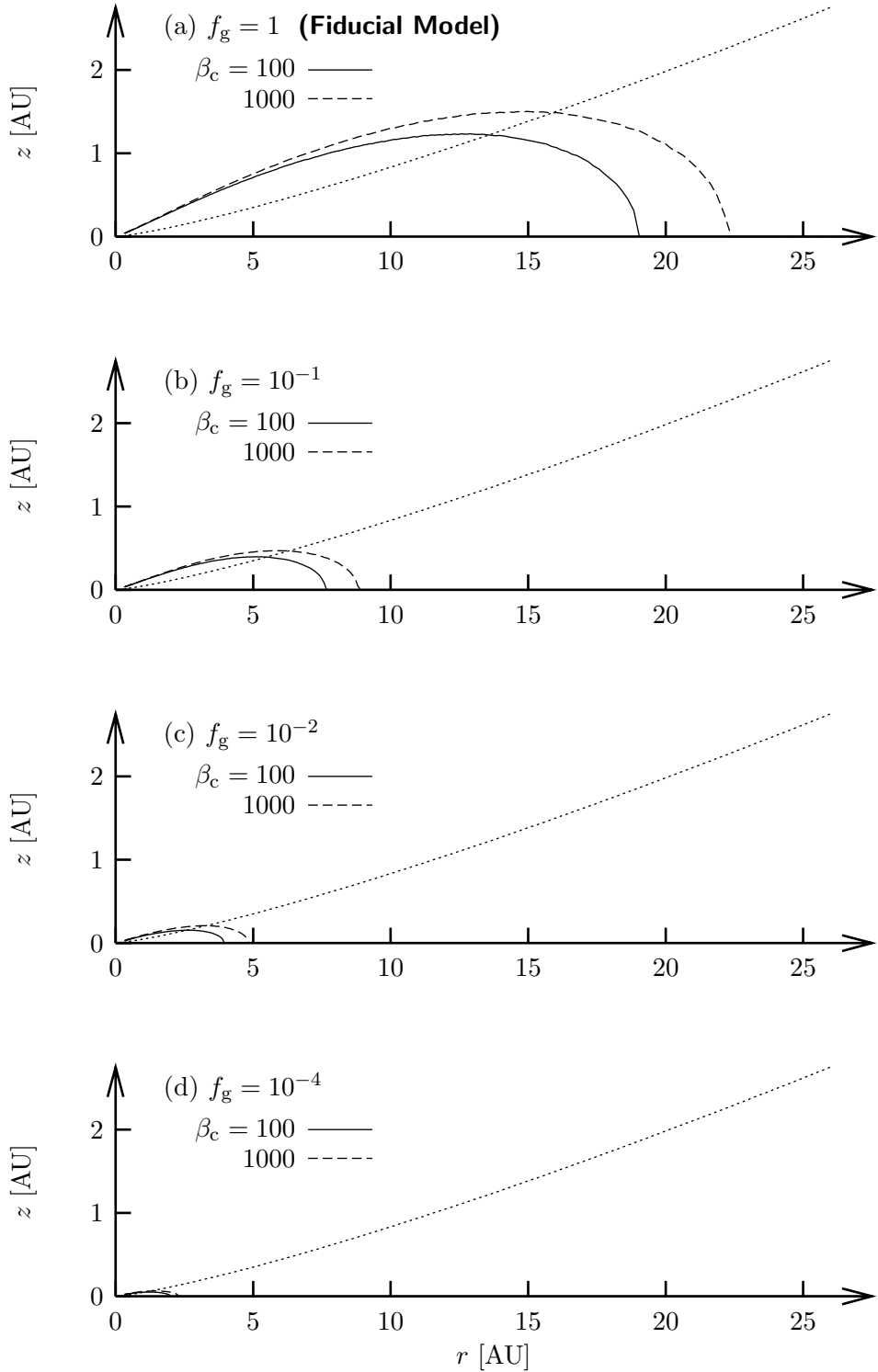
\begin{figure}
\begin{center}
\setlength{\unitlength}{0.1bp}
\begin{picture}(3600,1440)(0,0)
\put(1189,966){\makebox(0,0)[r]{1000}}
\put(1189,1084){\makebox(0,0)[r]{$\beta_{\rm c} = 100$}}
\put(647,1255){\makebox(0,0)[l]{(a) $f_{\rm g} = 1$ ~{\bsf (Fiducial Model)}}}
\put(2000,150){\makebox(0,0){ }}
\put(100,870){%
\makebox(0,0)[b]{\shortstack{$z$ [AU]}}%
}
\put(3268,300){\makebox(0,0){$25$}}
\put(2705,300){\makebox(0,0){$20$}}
\put(2141,300){\makebox(0,0){$15$}}
\put(1577,300){\makebox(0,0){$10$}}
\put(1014,300){\makebox(0,0){$5$}}
\put(450,300){\makebox(0,0){$0$}}
\put(400,1084){\makebox(0,0)[r]{$2$}}
\put(400,742){\makebox(0,0)[r]{$1$}}
\put(400,400){\makebox(0,0)[r]{$0$}}
\end{picture} \\
\setlength{\unitlength}{0.1bp}
\begin{picture}(3600,1440)(0,0)
\put(1189,966){\makebox(0,0)[r]{1000}}
\put(1189,1084){\makebox(0,0)[r]{$\beta_{\rm c} = 100$}}
\put(647,1255){\makebox(0,0)[l]{(b) $f_{\rm g} = 10^{-1}$}}
\put(2000,150){\makebox(0,0){ }}
\put(100,870){%
\makebox(0,0)[b]{\shortstack{$z$ [AU]}}%
}
\put(3268,300){\makebox(0,0){$25$}}
\put(2705,300){\makebox(0,0){$20$}}
\put(2141,300){\makebox(0,0){$15$}}
\put(1577,300){\makebox(0,0){$10$}}
\put(1014,300){\makebox(0,0){$5$}}
\put(450,300){\makebox(0,0){$0$}}
\put(400,1084){\makebox(0,0)[r]{$2$}}
\put(400,742){\makebox(0,0)[r]{$1$}}
\put(400,400){\makebox(0,0)[r]{$0$}}
\end{picture} \\
\setlength{\unitlength}{0.1bp}
\begin{picture}(3600,1440)(0,0)
\put(1189,966){\makebox(0,0)[r]{1000}}
\put(1189,1084){\makebox(0,0)[r]{$\beta_{\rm c} = 100$}}
\put(647,1255){\makebox(0,0)[l]{(c) $f_{\rm g} = 10^{-2}$}}
\put(2000,150){\makebox(0,0){ }}
\put(100,870){%
\makebox(0,0)[b]{\shortstack{$z$ [AU]}}%
}
\put(3268,300){\makebox(0,0){$25$}}
\put(2705,300){\makebox(0,0){$20$}}
\put(2141,300){\makebox(0,0){$15$}}
\put(1577,300){\makebox(0,0){$10$}}
\put(1014,300){\makebox(0,0){$5$}}
\put(450,300){\makebox(0,0){$0$}}
\put(400,1084){\makebox(0,0)[r]{$2$}}
\put(400,742){\makebox(0,0)[r]{$1$}}
\put(400,400){\makebox(0,0)[r]{$0$}}
\end{picture} \\
\setlength{\unitlength}{0.1bp}
\begin{picture}(3600,1440)(0,0)
\put(1189,966){\makebox(0,0)[r]{1000}}
\put(1189,1084){\makebox(0,0)[r]{$\beta_{\rm c} = 100$}}
\put(647,1255){\makebox(0,0)[l]{(d) $f_{\rm g} = 10^{-4}$}}
\put(2000,150){\makebox(0,0){$r$ [AU]}}
\put(100,870){%
\makebox(0,0)[b]{\shortstack{$z$ [AU]}}%
}
\put(3268,300){\makebox(0,0){$25$}}
\put(2705,300){\makebox(0,0){$20$}}
\put(2141,300){\makebox(0,0){$15$}}
\put(1577,300){\makebox(0,0){$10$}}
\put(1014,300){\makebox(0,0){$5$}}
\put(450,300){\makebox(0,0){$0$}}
\put(400,1084){\makebox(0,0)[r]{$2$}}
\put(400,742){\makebox(0,0)[r]{$1$}}
\put(400,400){\makebox(0,0)[r]{$0$}}
\end{picture} \\
\vspace{-12pt}
\caption
{Unstable regions for the disk of various evolutionary stages; 
(a) $f_{\rm g} = 1$, (b) $f_{\rm g} = 10^{-1}$,
(c) $f_{\rm g} = 10^{-2}$, and (d) $f_{\rm g} = 10^{-4}$. 
The other parameters are the same as those of the fiducial model.
The solid and dashed curves represent the loci of $\lambda_{\rm res} / H
= 1$ for the cases of the field strength $\beta_{\rm c} = 100$ and
1000, respectively, inside of which is magnetorotationally stable.
The dotted curve represents the scale height of the disk $z = H (r)$.
\label{fig:fg}}
\end{center}
\end{figure}

To clarify the effect of grain depletion, we investigate several cases
with respect to the depletion factor of dust grains, (a) $f_{\rm g} =
1$ (fiducial model), (b) $f_{\rm g} = 10^{-1}$, (c) $f_{\rm g} =
10^{-2}$, and (d) $f_{\rm g} = 10^{-4}$, each of which might
correspond to some evolutionary stage of the protoplanetary disk.  
Figure~\ref{fig:fg} shows the results for the disk with the surface
density parameter $f_{\Sigma} = 1$ around a star of $M_{\ast} = 1
~M_{\odot}$.
The other parameters are the same as in the fiducial model.

We find that decrease of dust abundance makes smaller the region where
magnetorotational instability is suppressed by ohmic dissipation.
At later stages where only a very small fraction of the initial dust
grains floats, magnetic fields are nearly frozen to the gas in most
parts of the disk.
This implies that the magnetorotational instability is operative at a
larger part of the disk and the angular momentum transport is more
effective in the later evolutionary stages. 
For the case of $\beta_{\rm c} = 100$, the critical radius $r_{\rm
crit}$ is 19, 7.6, 3.8, and 2.0 AU for the models of $f_{\rm g} = 1$,
$10^{-1}$, $10^{-2}$, and $10^{-4}$, respectively.

For $f_{\rm g} = 10^{-4}$, the critical height $z_{\rm crit}$ where
$\lambda_{\rm res} / H = 1$ holds is $1.0 ~H$ and $1.2 ~H$ for the
cases of $\beta_{\rm c} = 100$ and 1000, respectively, at 1 AU.
At these points, electrons and metal ions are still dominant charged
particles (see Fig.~\ref{fig:zdep-fg}). 
At $|z| = z_{\rm crit}$ in the region of $r \approx 1$ AU, the
ionization rate is smaller than $\zeta_{\rm CR}$ by an order of magnitude. 
This is because the column density above the dead zone is nearly equal
to $\chi_{\rm CR} \approx 96$ g cm$^{-2}$ for these cases.

\subsubsection{Growth of Grains}

\begin{figure}
\begin{center}
\setlength{\unitlength}{0.1bp}
\begin{picture}(3600,1440)(0,0)
\put(1189,966){\makebox(0,0)[r]{1000}}
\put(1189,1084){\makebox(0,0)[r]{$\beta_{\rm c} = 100$}}
\put(647,1255){\makebox(0,0)[l]{(a) $a = 0.03$ $\mu$m}}
\put(2000,150){\makebox(0,0){ }}
\put(100,870){%
\makebox(0,0)[b]{\shortstack{$z$ [AU]}}%
}
\put(3268,300){\makebox(0,0){$50$}}
\put(2705,300){\makebox(0,0){$40$}}
\put(2141,300){\makebox(0,0){$30$}}
\put(1577,300){\makebox(0,0){$20$}}
\put(1014,300){\makebox(0,0){$10$}}
\put(450,300){\makebox(0,0){$0$}}
\put(400,1255){\makebox(0,0)[r]{$5$}}
\put(400,1084){\makebox(0,0)[r]{$4$}}
\put(400,913){\makebox(0,0)[r]{$3$}}
\put(400,742){\makebox(0,0)[r]{$2$}}
\put(400,571){\makebox(0,0)[r]{$1$}}
\put(400,400){\makebox(0,0)[r]{$0$}}
\end{picture} \\
\setlength{\unitlength}{0.1bp}
\begin{picture}(3600,1440)(0,0)
\put(1189,966){\makebox(0,0)[r]{1000}}
\put(1189,1084){\makebox(0,0)[r]{$\beta_{\rm c} = 100$}}
\put(647,1255){\makebox(0,0)[l]{(b) $a = 0.1$ $\mu$m ~{\bsf (Fiducial Model)}}}
\put(2000,150){\makebox(0,0){ }}
\put(100,870){%
\makebox(0,0)[b]{\shortstack{$z$ [AU]}}%
}
\put(3268,300){\makebox(0,0){$50$}}
\put(2705,300){\makebox(0,0){$40$}}
\put(2141,300){\makebox(0,0){$30$}}
\put(1577,300){\makebox(0,0){$20$}}
\put(1014,300){\makebox(0,0){$10$}}
\put(450,300){\makebox(0,0){$0$}}
\put(400,1255){\makebox(0,0)[r]{$5$}}
\put(400,1084){\makebox(0,0)[r]{$4$}}
\put(400,913){\makebox(0,0)[r]{$3$}}
\put(400,742){\makebox(0,0)[r]{$2$}}
\put(400,571){\makebox(0,0)[r]{$1$}}
\put(400,400){\makebox(0,0)[r]{$0$}}
\end{picture} \\
\setlength{\unitlength}{0.1bp}
\begin{picture}(3600,1440)(0,0)
\put(1189,966){\makebox(0,0)[r]{1000}}
\put(1189,1084){\makebox(0,0)[r]{$\beta_{\rm c} = 100$}}
\put(647,1255){\makebox(0,0)[l]{(c) $a = 0.3$ $\mu$m}}
\put(2000,150){\makebox(0,0){ }}
\put(100,870){%
\makebox(0,0)[b]{\shortstack{$z$ [AU]}}%
}
\put(3268,300){\makebox(0,0){$50$}}
\put(2705,300){\makebox(0,0){$40$}}
\put(2141,300){\makebox(0,0){$30$}}
\put(1577,300){\makebox(0,0){$20$}}
\put(1014,300){\makebox(0,0){$10$}}
\put(450,300){\makebox(0,0){$0$}}
\put(400,1255){\makebox(0,0)[r]{$5$}}
\put(400,1084){\makebox(0,0)[r]{$4$}}
\put(400,913){\makebox(0,0)[r]{$3$}}
\put(400,742){\makebox(0,0)[r]{$2$}}
\put(400,571){\makebox(0,0)[r]{$1$}}
\put(400,400){\makebox(0,0)[r]{$0$}}
\end{picture} \\
\setlength{\unitlength}{0.1bp}
\begin{picture}(3600,1440)(0,0)
\put(1189,966){\makebox(0,0)[r]{1000}}
\put(1189,1084){\makebox(0,0)[r]{$\beta_{\rm c} = 100$}}
\put(647,1255){\makebox(0,0)[l]{(d) $a = 1$ $\mu$m}}
\put(2000,150){\makebox(0,0){$r$ [AU]}}
\put(100,870){%
\makebox(0,0)[b]{\shortstack{$z$ [AU]}}%
}
\put(3268,300){\makebox(0,0){$50$}}
\put(2705,300){\makebox(0,0){$40$}}
\put(2141,300){\makebox(0,0){$30$}}
\put(1577,300){\makebox(0,0){$20$}}
\put(1014,300){\makebox(0,0){$10$}}
\put(450,300){\makebox(0,0){$0$}}
\put(400,1255){\makebox(0,0)[r]{$5$}}
\put(400,1084){\makebox(0,0)[r]{$4$}}
\put(400,913){\makebox(0,0)[r]{$3$}}
\put(400,742){\makebox(0,0)[r]{$2$}}
\put(400,571){\makebox(0,0)[r]{$1$}}
\put(400,400){\makebox(0,0)[r]{$0$}}
\end{picture} \\
\vspace{-12pt}
\caption
{Unstable regions for the disks with various grain sizes; 
(a) $a = 0.03$ $\mu$m, (b) $a = 0.1$ $\mu$m, (c) $a = 0.3$ $\mu$m, and
(d) $a = 1$ $\mu$m.
The other parameters are the same as those of the fiducial model.
The solid and dashed curves represent the loci of $\lambda_{\rm res} / H
= 1$ for the cases of the field strength $\beta_{\rm c} = 100$ and
1000, respectively, inside of which is magnetorotationally stable.
The dotted curve represents the scale height of the disk $z = H (r)$.
\label{fig:a}}
\end{center}
\end{figure}
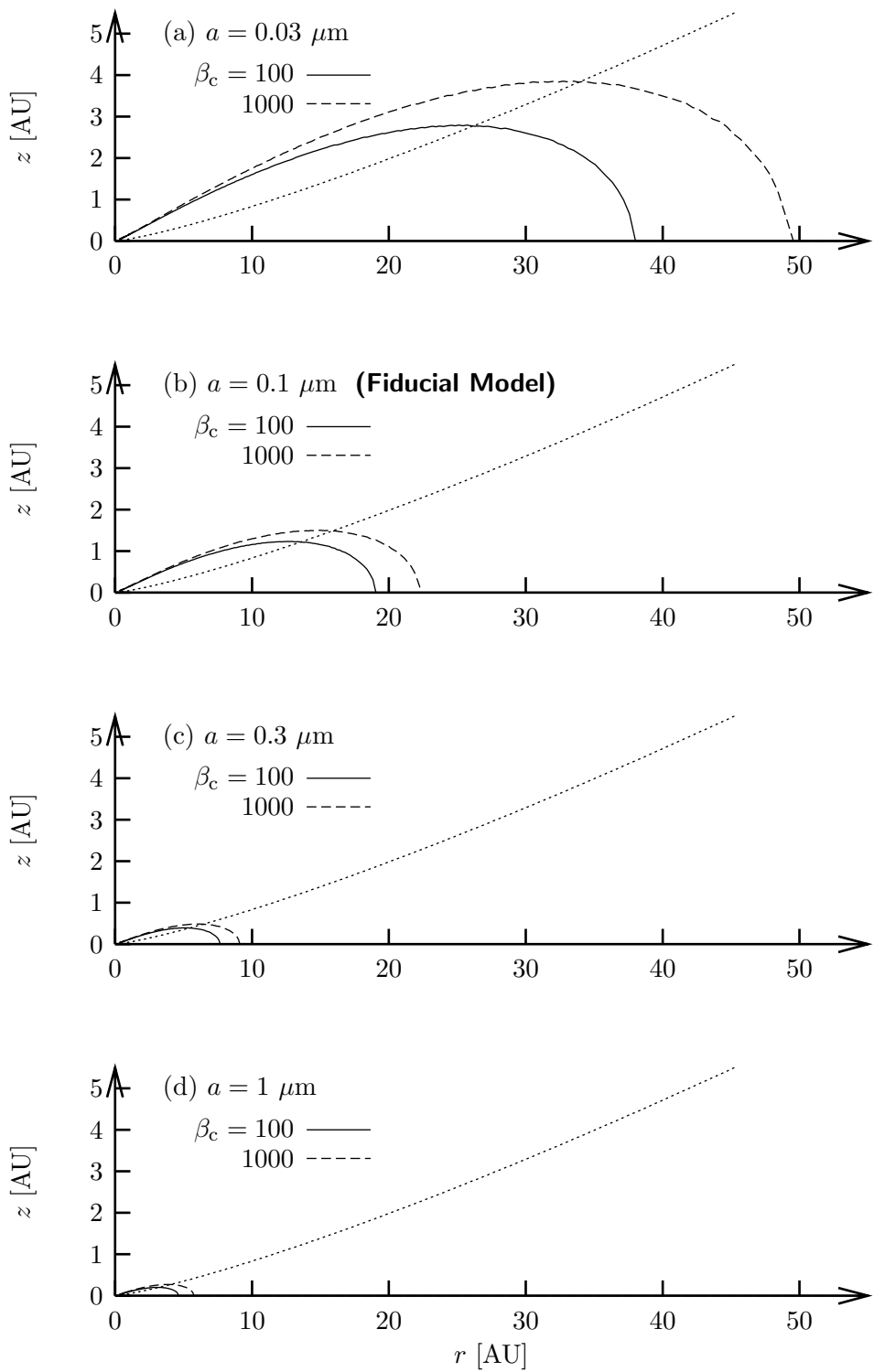

The interstellar grains have a wide size distribution, e.g., between
0.005 and 0.25 $\mu$m (Mathis, Rumpl, \& Nordsieck 1977; hereafter
MRN).
By coalescence of dust grains in protoplanetary disks, the size
distribution and the relative number density of grains may change
considerably.  
To clarify the effect of grain size on the unstable regions, we
investigate the following four cases assuming for simplicity that all
grains have the same radius; (a) $a = 0.03$ $\mu$m, (b) $a = 0.1$
$\mu$m (fiducial model), (c) $a = 0.3$ $\mu$m, and (d) $a = 1$ $\mu$m.
The relative number density of dust grains $n_{\rm g}/ n_{\rm H}$ is
proportional to $a^{-3}$ and is, for example, $3.3 \times 10^{-11}$
and $8.9 \times 10^{-13}$ for $a = 0.03$ and $0.1$ $\mu$m,
respectively.
The results for a disk of $f_{\Sigma} = 1$ and $f_{\rm g} = 1$ around
a star of $M_{\ast} = 1 ~M_{\odot}$ are shown in Figure~\ref{fig:a}.  
The other parameters are the same as in the fiducial model.

We find that the dead zone shrinks as the grain size increases.
As the grains grow, the ion density increases because the total surface
area of grains decreases and thus the recombination rate on grain
surface decreases.
For the case of $\beta_{\rm c} = 100$, the critical radius $r_{\rm
crit}$ is 37, 19, 7.6, and 4.6 AU for the models with $a = 0.03$, 0.1, 
0.3, and 1 $\mu$m, respectively.

For the case of $a = 0.03$ $\mu$m, the number density of dust grains
becomes comparable to or higher than that of metal ions at $n_{\rm H}
\gtrsim 10^8 ~{\rm cm}^{-3}$.
At the critical radius $r_{\rm crit} \approx 37$ AU on the midplane,
the abundance of metal ions $x ({\rm M}^{+}) \approx 1.4 \times
10^{-12}$ is comparable to those of charged grains with $\pm e$,
though the conductivity $\sigma_{\rm c}$ is still mainly contributed by
electrons with abundance $x ({\rm e}) \approx 0.9 \times 10^{-14}$.

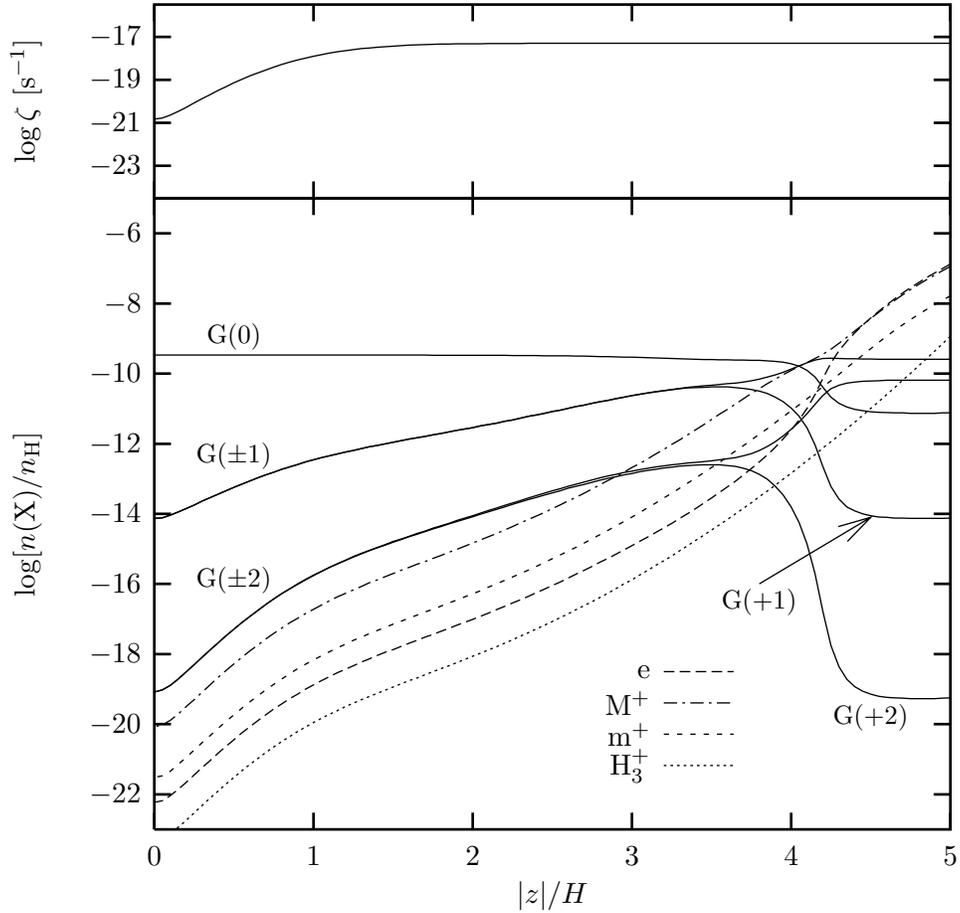
\begin{figure}
\begin{center}
\setlength{\unitlength}{0.1bp}
\begin{picture}(3600,1080)(0,0)
\put(100,615){%
\makebox(0,0)[b]{\shortstack{$\log \zeta ~[ {\rm s}^{-1} ]$}}%
}
\put(500,858){\makebox(0,0)[r]{$-17$}}
\put(500,696){\makebox(0,0)[r]{$-19$}}
\put(500,534){\makebox(0,0)[r]{$-21$}}
\put(500,372){\makebox(0,0)[r]{$-23$}}
\end{picture} \\
\vspace{-36pt}
\setlength{\unitlength}{0.1bp}
\begin{picture}(3600,2880)(0,0)
\put(2420,641){\makebox(0,0)[r]{H$_3^{+}$}}
\put(2420,759){\makebox(0,0)[r]{m$^{+}$}}
\put(2420,877){\makebox(0,0)[r]{M$^{+}$}}
\put(2420,995){\makebox(0,0)[r]{e}}
\put(3250,823){\makebox(0,0){{\small G($+$2)}}}
\put(2830,1259){\makebox(0,0){{\small G($+$1)}}}
\put(850,1339){\makebox(0,0){{\small G($\pm$2)}}}
\put(850,1808){\makebox(0,0){{\small G($\pm$1)}}}
\put(850,2258){\makebox(0,0){{\small G(0)}}}
\put(2050,150){\makebox(0,0){$|z| / H$}}
\put(100,1590){%
\makebox(0,0)[b]{\shortstack{$\log [ n({\rm X}) / n_{\rm H} ]$}}%
}
\put(3550,300){\makebox(0,0){$5$}}
\put(2950,300){\makebox(0,0){$4$}}
\put(2350,300){\makebox(0,0){$3$}}
\put(1750,300){\makebox(0,0){$2$}}
\put(1150,300){\makebox(0,0){$1$}}
\put(550,300){\makebox(0,0){$0$}}
\put(500,2648){\makebox(0,0)[r]{$-6$}}
\put(500,2383){\makebox(0,0)[r]{$-8$}}
\put(500,2119){\makebox(0,0)[r]{$-10$}}
\put(500,1854){\makebox(0,0)[r]{$-12$}}
\put(500,1590){\makebox(0,0)[r]{$-14$}}
\put(500,1326){\makebox(0,0)[r]{$-16$}}
\put(500,1061){\makebox(0,0)[r]{$-18$}}
\put(500,797){\makebox(0,0)[r]{$-20$}}
\put(500,532){\makebox(0,0)[r]{$-22$}}
\end{picture} \\
\vspace{-12pt}
\caption
{Relative abundances $n({\rm X})/n_{\rm H}$ of some representative
particles as functions of the height $|z|$ from the midplane at $r =
1$ AU for the MRN size distribution of grains.
The other parameters are the same as those of the fiducial model.
The upper panel shows the ionization rate $\zeta$.
\label{fig:zdep-mrn}}
\end{center}
\end{figure}

So far we have assumed that all grains have the same radius.  
However, interstellar grains have a wide size distribution such as the
MRN distribution for graphite and silicate grains,
\begin{equation}
\frac{d n_{\rm g}}{d a} = A \, n_{\rm H} \, a^{-3.5} \;, \quad
a_{\min} < a < a_{\max} \;,
\label{eqn:mrn}
\end{equation}
where $a_{\min} \approx 0.005$ $\mu$m, $a_{\max} \approx 0.25$ $\mu$m,
and $A \approx 1.5 \times 10^{-25}$ cm$^{2.5}$ (Draine \& Lee 1984).
For comparison, we show the results for this size distribution.
Figure~\ref{fig:zdep-mrn} shows the vertical distribution of the
relative abundances $n({\rm X})/n_{\rm H}$ of some representative
particles at 1 AU.

The qualitative features are essentially the same as those for the
fiducial model (see Fig.~\ref{fig:zdep}).
The total surface area of grains is contributed mainly by grains with
radius $a \approx a_{\min}$ in the MRN distribution.  
Therefore, decrease of $a_{\min}$ enhances the recombination rate of
ions and electrons on grains; $a_{\min}$ can be smaller than 0.005
$\mu$m (MRN).
Dust grains with electric charge $\pm e$ are the dominant charged
particles at the midplane everywhere within $r \approx$ 100 AU.
In this case the magnetic field dissipation is more effective than in
the fiducial model. 
The magnetorotational instability is suppressed in a wider region, and
the critical radii on the midplane $r_{\rm crit} = 44$ and 59 AU for
the cases of $\beta_{\rm c} = 100$ and 1000, respectively, are 2 to 3
times larger than those for the fiducial model.

\section{DISCUSSION}

\subsection{Layered Accretion}

In paper I we have found that, in some disk conditions, the unstable
modes have large amplitudes localized at the upper layer of the disk.
This implies that the magnetorotational instability in the nonlinear
regime enables the angular momentum transport only in this layer, and
thus the layered accretion occurs as proposed by Gammie (1996).
This layer corresponds to the unstable region at $r \lesssim r_{\rm
crit}$ shown by the stripes in Figure~\ref{fig:d100}.
Since this unstable layer locates at a height a few times the scale
height of the disk, the column density of this layer is very small.  
Therefore, the layered accretion might hardly affect dynamical
evolution of the disks.
However, the thickness of this layer is determined by the distribution
of the magnetic diffusivity, which depends not only on the density
distribution but also on the size and abundance of grains.  

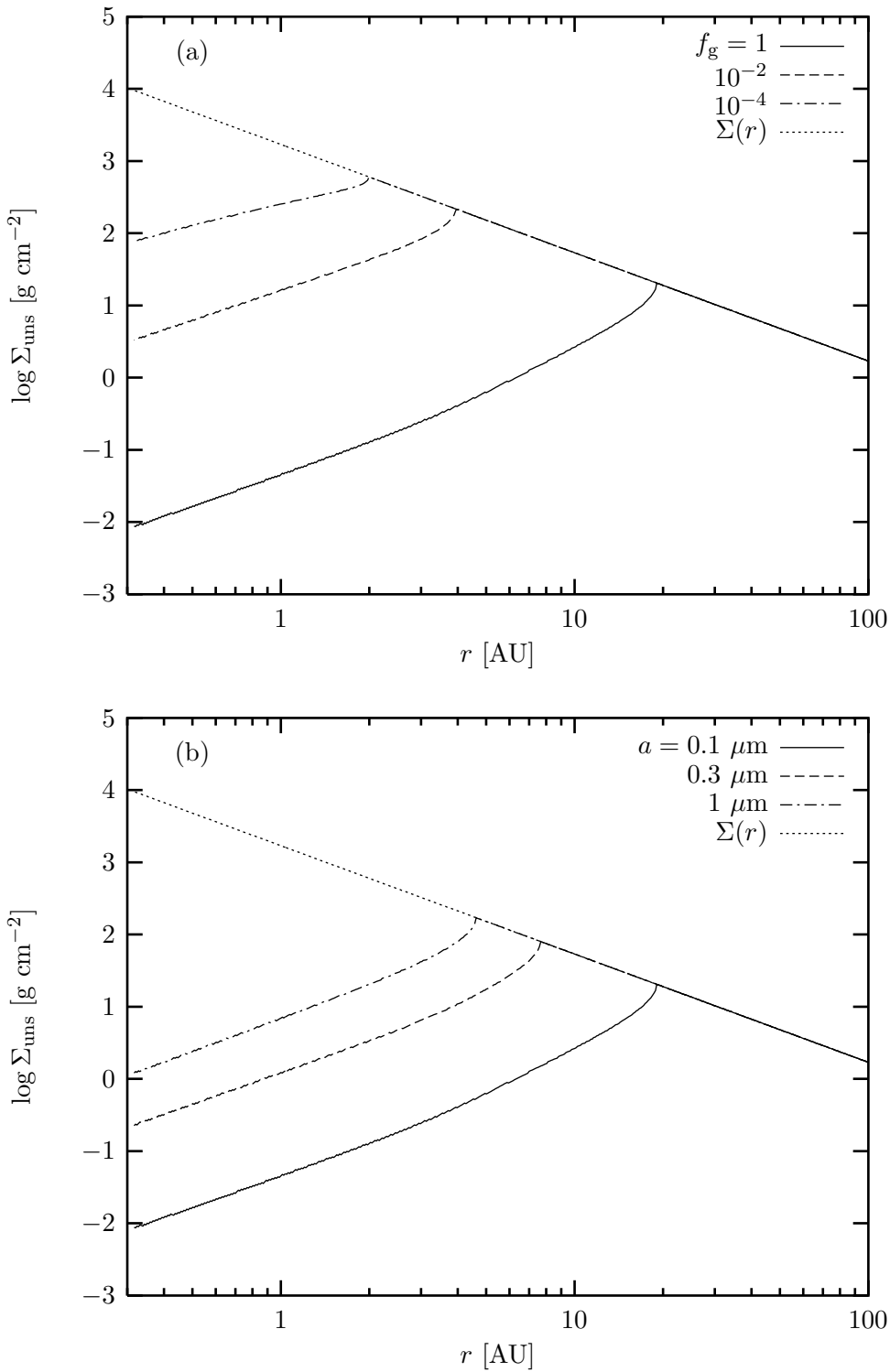
\begin{figure}
\begin{center}
\setlength{\unitlength}{0.1bp}
\begin{picture}(3600,2880)(0,0)
\put(3137,2304){\makebox(0,0)[r]{$\Sigma(r)$}}
\put(3137,2422){\makebox(0,0)[r]{$10^{-4}$}}
\put(3137,2540){\makebox(0,0)[r]{$10^{-2}$}}
\put(3137,2658){\makebox(0,0)[r]{$f_{\rm g} = 1$}}
\put(768,2631){\makebox(0,0){(a)}}
\put(2025,150){\makebox(0,0){$r$ [AU]}}
\put(100,1590){%
\makebox(0,0)[b]{\shortstack{$\log \Sigma_{\rm uns}$ [g~cm$^{-2}$]}}%
}
\put(3550,300){\makebox(0,0){$100$}}
\put(2341,300){\makebox(0,0){$10$}}
\put(1132,300){\makebox(0,0){$1$}}
\put(450,2780){\makebox(0,0)[r]{$5$}}
\put(450,2483){\makebox(0,0)[r]{$4$}}
\put(450,2185){\makebox(0,0)[r]{$3$}}
\put(450,1888){\makebox(0,0)[r]{$2$}}
\put(450,1590){\makebox(0,0)[r]{$1$}}
\put(450,1293){\makebox(0,0)[r]{$0$}}
\put(450,995){\makebox(0,0)[r]{$-1$}}
\put(450,698){\makebox(0,0)[r]{$-2$}}
\put(450,400){\makebox(0,0)[r]{$-3$}}
\end{picture} \\
\setlength{\unitlength}{0.1bp}
\begin{picture}(3600,2880)(0,0)
\put(3137,2304){\makebox(0,0)[r]{$\Sigma(r)$}}
\put(3137,2422){\makebox(0,0)[r]{$1$ $\mu$m}}
\put(3137,2540){\makebox(0,0)[r]{$0.3$ $\mu$m}}
\put(3137,2658){\makebox(0,0)[r]{$a = 0.1$ $\mu$m}}
\put(768,2631){\makebox(0,0){(b)}}
\put(2025,150){\makebox(0,0){$r$ [AU]}}
\put(100,1590){%
\makebox(0,0)[b]{\shortstack{$\log \Sigma_{\rm uns}$ [g~cm$^{-2}$]}}%
}
\put(3550,300){\makebox(0,0){$100$}}
\put(2341,300){\makebox(0,0){$10$}}
\put(1132,300){\makebox(0,0){$1$}}
\put(450,2780){\makebox(0,0)[r]{$5$}}
\put(450,2483){\makebox(0,0)[r]{$4$}}
\put(450,2185){\makebox(0,0)[r]{$3$}}
\put(450,1888){\makebox(0,0)[r]{$2$}}
\put(450,1590){\makebox(0,0)[r]{$1$}}
\put(450,1293){\makebox(0,0)[r]{$0$}}
\put(450,995){\makebox(0,0)[r]{$-1$}}
\put(450,698){\makebox(0,0)[r]{$-2$}}
\put(450,400){\makebox(0,0)[r]{$-3$}}
\end{picture} \\
\vspace{-12pt}
\caption
{The column density of the unstable layer, $\Sigma_{\rm uns}$,
(a) for the models of grain depletion $f_{\rm g} = 1$, $10^{-2}$, and
$10^{-4}$, and (b) for the models of grain radius
$a = 0.1$, 0.3, and 1 $\mu$m.
The other parameters are the same as those of the fiducial model. 
\label{fig:col2}}
\end{center}
\end{figure}

First, we examine the dependence of the column density of the unstable
layer, $\Sigma_{\rm uns}$, on the depletion factor of dust grains,
$f_{\rm g}$.
Because the column density of the layer above the critical curve
$\lambda_{\rm ideal} / H = 1$ is quite small, we approximate
$\Sigma_{\rm uns}$ as the total column density of the disk subtracted
by the column density of the dead zone. 
Figure~\ref{fig:col2}a shows $\Sigma_{\rm uns}$ as a function of $r$
for $\beta = 100$ for three cases of grain depletion, $f_{\rm g} = 1$,
$10^{-2}$, and $10^{-4}$.
The grain size is taken to be $a = 0.1$ $\mu$m and the other parameters
are the same as in the fiducial model.
The column density $\Sigma_{\rm uns}$ increases as the grain fraction
decreases.
Gammie (1996) assumed that the column density of the unstable region
was comparable to the attenuation length $\chi_{\rm CR} \approx 96$ g
cm$^{-2}$.
We have found that the column density $\Sigma_{\rm uns}$ is much
smaller than $\chi_{\rm CR}$ unless $f_{\rm g} \lesssim 10^{-4}$.
Magnetic fields can be decoupled from the gas even when cosmic rays
are not attenuated.

Next we examine the dependence of $\Sigma_{\rm uns}$ on the grain
size.
Figure~\ref{fig:col2}b shows the column density $\Sigma_{\rm uns}$ for
the models with $a = 0.1$, 0.3, and 1 $\mu$m.
Here we have assumed the grain fraction $f_{\rm g} = 1$ and
$\beta_{\rm c} = 100$.
The column density $\Sigma_{\rm uns}$ increases as the grain radius
increases.
However, $\Sigma_{\rm uns}$ is at most a few g cm$^{-2}$ within $r
\sim 1$ AU, even when the grain is as large as $a = 1$ $\mu$m.
Therefore, unless the abundance of dust grains is reduced to $10^{-4}$
times the interstellar abundance, mass accretion through the
unstable layer above the dead zone is very faint and the matter must
accumulate around the outer boundary of the dead zone.

\subsection{Mass Accretion Rate}

We estimate the mass accumulation rate into the dead zone.
Sano, Inutsuka, \& Miyama (1998) investigated the nonlinear evolution
of magnetorotational instability including the effect of ohmic
dissipation using two-dimensional MHD simulations.
Their results show that, at the saturated turbulent state, the
efficiency of angular momentum transport depends on the strength of
the initial magnetic field perpendicular to the disk, and that when the
magnetic Reynolds number $R_{\rm m} \equiv v_{\rm A}^2 / \eta \Omega$
is around unity, the $\alpha$ parameter of the viscosity (Shakura \&
Sunyaev 1973) is given by 
\begin{equation}
\alpha \approx 1.8 \times 10^{-2} \left( \frac{\beta}{1000}
\right)^{-1} \;,
\label{eqn:alpha}
\end{equation}
where $\beta$ is the plasma beta for the initial field.

Because the magnetic Reynolds number $R_{\rm m}$ is close to unity
around the critical radius, the mass accretion rate can be estimated
with equation~(\ref{eqn:alpha}).
The radial accretion velocity $v_r$ is given by $v_r \sim \nu / r$,
where $\nu = \alpha c_{\rm s} H$ is the kinematic viscosity. 
Then the mass accretion rate at $r \approx r_{\rm crit}$ is given by
\begin{equation}
\dot{M} 
\sim 2 \pi r_{\rm crit} \Sigma v_r 
\approx 4.9 \times 10^{-7} f_{\Sigma}
\left( \frac{r_{\rm crit}}{19 ~{\rm AU}} \right)^{-1/2}
\left( \frac{\beta_{\rm c}}{100} \right)^{-1} \quad 
M_{\odot}~{\rm yr}^{-1} \;,
\label{eqn:mdot}
\end{equation}
for the disks whose structure is given by equations~(\ref{eqn:sigr}),
(\ref{eqn:csr}), and (\ref{eqn:hr}) with $q = 3 / 2$ and $M_{\ast} = 1
~M_{\odot}$.
For the case of $\beta_{\rm c} = 1000$, the mass accretion rate at $r
\approx r_{\rm crit} \approx 22$ AU is an order of magnitude smaller.

The typical age of T Tauri stars is estimated to be $10^6$ yr
(Kenyon \& Hartmann 1995), and the lifetime of protoplanetary disks
must be comparable to or longer than the age of T Tauri stars.
If mass accretion with a rate $\dot{M} \sim 10^{-6}$
$M_{\odot}$~yr$^{-1}$ continues and the accreted matter is distributed 
relatively widely inside $r \approx r_{\rm crit}$, the disk of this
part becomes gravitationally unstable at some stage, and the resultant 
gravitational torque causes mass accretion in this region.
This makes the accretion onto the central star non-steady.
If the accreting matter accumulates near the critical radius $r
\approx r_{\rm crit}$ and a dense ring forms, a planet or a brown
dwarf would be formed due to fragmentation of the ring (Nakano 1991).
In this case, the critical radius gives the position where the first
small companion forms.

Higher mass disks have higher accretion rate as seen from
equation~(\ref{eqn:mdot}).
The mass accretion rate for a disk of $f_{\Sigma} = 10$ at $r \approx
r_{\rm crit} \approx 39$ AU is 7 times larger than that of the
fiducial model at $r_{\rm crit} \approx 19$ AU.  
For the model of $f_{\Sigma} = 0.3$, we have the mass accretion rate
$\dot{M} \sim 1.8 \times 10^{-7} ~M_{\odot}~{\rm yr}^{-1}$ at $r
\approx r_{\rm crit} \approx 13$ AU.   
Therefore, higher mass disks become gravitationally unstable in
shorter time scale.

\subsection{Evolution of Dust Grains in Protoplanetary Disks}

Investigating the growth and sedimentation of dust grains in
protoplanetary disks, Nakagawa et al. (1981) showed that dust grains
can grow by sticking through mutual collisions in thermal Brownian
motion, e.g., when the grain size is smaller than microns at $r
\approx 1$ AU.  
The collision time in Brownian motion, $t_{\rm col}$, is given by 
\begin{equation}
t_{\rm col} = 2 \pi \frac{(\rho_{\rm g}/3)^{3/2} \,
\bar{a}^{5/2}}{\zeta_{\rm g} \, \rho \, (kT)^{1/2}}
= 21 
\left( \frac{\rho_{\rm g}}{3 ~{\rm g}~{\rm cm}^{-3}} \right)^{3/2}
\left( \frac{\bar{a}}{1 ~\mu {\rm m}} \right)^{5/2} 
\left( \frac{\zeta_{\rm g}}{0.0034} \right)^{-1} 
\left( \frac{r}{1 ~{\rm AU}} \right)^{3/2}
t_{\rm K} \;,
\label{eqn:growth}
\end{equation}
where $\rho_{\rm g}$ and $\bar{a}$ are the internal density and the
mean radius, respectively, of dust grains, and $\zeta_{\rm g}$ is the
mass fraction of dust grains, and the last expression is at the
midplane of the fiducial model.
As seen from equation~(\ref{eqn:growth}), dust grains in dense regions
grow to micron size in a time comparable to the dynamical time scale,
$t_{\rm K}$, and much shorter than the time scale of sedimentation,
\begin{equation}
t_{\rm sed} =
\frac1{2 \pi} \frac{\rho H}{\rho_{\rm g} \bar{a}} t_{\rm K} 
= 5.2 \times 10^{5} 
\left( \frac{\rho_{\rm g}}{3 ~{\rm g}~{\rm cm}^{-3}} \right)^{-1}
\left( \frac{\bar{a}}{1 ~\mu {\rm m}} \right)^{-1} 
 {\rm yr} \;,
\end{equation}
where the last expression is at the midplane of the fiducial model and is 
independent of $r$.

Numerical calculations of Nakagawa et al. (1981) show that the mass
fraction of floating grains, $f_{\rm g}$, at $r = 1$ AU is $\sim
10^{-1}$, $10^{-2}$, and $10^{-4}$ at the time $t \sim 2 \times 10^3$,
$10^4$, and $10^5$ yr, respectively, and that the size of floating
grains, $a$, is 1 -- 10 $\mu$m in the late stages.
Although laminar flow was assumed in their analysis, we discuss the
evolution of the unstable regions using their results.
Figure~\ref{fig:a}d shows the unstable region for the model with $a = 1$
$\mu$m, an order of magnitude larger than in the fiducial model.
The critical radius on the midplane $r_{\rm crit}$ is 4.6 and 5.8 AU for 
the cases of $\beta_{\rm c} = 100$ and $1000$, respectively.
The dead zone shrinks as the sedimentation of dust grains proceeds; in
the cases of $f_{\rm g} = 10^{-1}$, $f_{\rm g} = 10^{-2}$, and $f_{\rm
g} = 10^{-4}$, the critical radius of the models with $a = 1$ $\mu$m is
3.2 (3.8), 2.3 (2.7), and 1.4 (1.6) AU, respectively, for $\beta_{\rm c}
= 100$ (1000).
In all these cases the critical radius $r_{\rm crit}$ is less than 6 AU. 
If dust grains in protoplanetary disks could grow to micron size,
most parts of the disk become magnetorotationally unstable.
Therefore, sedimentation of dust grains must be prevented by turbulence
induced by this instability contrary to the assumption of Nakagawa et
al. (1981).

According to the standard scenario of solar system formation (Hayashi
et al. 1985), planets form by accretion of planetesimals, which are
produced through gravitational instability of a thin dust layer
(Goldreich \& Ward 1973).  
The thin layer is formed by sedimentation of dust, which must occur only 
in a quiescent disk (Mizuno, Markiewicz, \& V{\"o}lk 1988).
By sedimentation the disk becomes magnetorotationally unstable and
turbulent, preventing sedimentation as shown in \S 3.3.
Therefore we have to investigate the gas motion and the evolution of
dust simultaneously.

\subsection{Ionization by X-Rays}

So far we have considered only cosmic rays and radioactive elements as
the ionization sources of protoplanetary disks.
In this subsection we consider the effect of photo-processes.
The protoplanetary disks are exposed to the UV radiation from the
central star and from the interstellar space.
Because of the very small attenuation length, a few $\times$ 10$^{-3}$ g
cm$^{-2}$, the ionization by UV radiation is inefficient except for very 
thin surface layers.

Young stellar objects are strong X-ray sources (e.g., Glassgold,
Feigelson, \& Montmerle 2000).
The X-ray luminosity of low-mass young stellar objects is typically in
the range $L_{\rm X} \sim 10^{28}$ -- $10^{30}$ erg s$^{-1}$, or a
factor of $10^2$ -- $10^3$ above the contemporary solar levels.
Some X-ray sources exhibit high-amplitude rapid flares with peak
luminosity $L_{\rm X} \sim 10^{30}$ -- $10^{32}$ erg s$^{-1}$.
These emissions require the presence of a large volume of high-density
plasma at a temperature of about $10^{7}$ K. 
The plasma must be magnetically confined, probably in large loops on a 
scale comparable to or larger than the X-ray emitting star itself.
Field lines linking the star with the disk at the corotating radius
may explain the strongest of these powerful flares (Shu et al. 1994;
Hayashi, Shibata, \& Matsumoto 1996).
The X-rays can induce a wide variety of changes in chemical and
physical properties of protoplanetary disks.

Soft X-rays can be the most promising ionization sources for
protoplanetary disks as well as cosmic rays.
While the underlying ionization mechanism is due to electronic
collisions in both cases, X-rays are absorbed in a smaller column of
matter than cosmic rays.
Thus cosmic rays provide a more global, low-level ionization of the
disks, whereas X-rays can produce a localized high-level ionization.

Igea \& Glassgold (1999) investigated the X-ray transfer and ionization
in the disks around young stellar objects using a Monte Carlo method.
The X-ray emitting region was modeled as a ring of radius $r = 10
~R_{\odot}$ at height $z = 10 ~R_{\odot}$ with $L_{\rm X} = 10^{29}$
erg s$^{-1}$ based on the x-wind model of Shu et al. (1994).
They obtained the ionization rates at 1 -- 10 AU of the minimum-mass
solar nebula as functions of the vertical column density into the disk 
by hydrogen number, $N_{\perp}$.
Their results show that the ionization rate by X-rays is larger than
that by cosmic rays, $\zeta_{\rm CR} \approx 10^{-17}$ s$^{-1}$, only
at $N_{\perp}$ smaller than $10^{25}$ to $2 \times 10^{23}$ cm$^{-2}$
depending on $r$, which is much smaller than the total column density
of the disk at each $r$.
Therefore, most of our results are preserved even if X-ray ionization is
taken into account.
For instance, the thickness of the dead zone is hardly affected by the
X-ray ionization. 
Even if cosmic rays are excluded from the inner regions of
protoplanetary disks by, e.g., stellar winds (Parker 1960), the active 
surface layer can be maintained by the X-ray ionization (Igea \&
Glassgold 1999).
Igea \& Glassgold (1999) neglected the effects of dust grains on ion
densities and on interaction with magnetic fields.
However, the column densities of unstable regions $\Sigma_{\rm uns}$
must be affected by the characters of dust grains in the disk in the
same way as in the cases with the cosmic ray ionization investigated in
\S 3.  

\subsection{Uncertainties in the Disk Temperature}
\label{sec:temp}

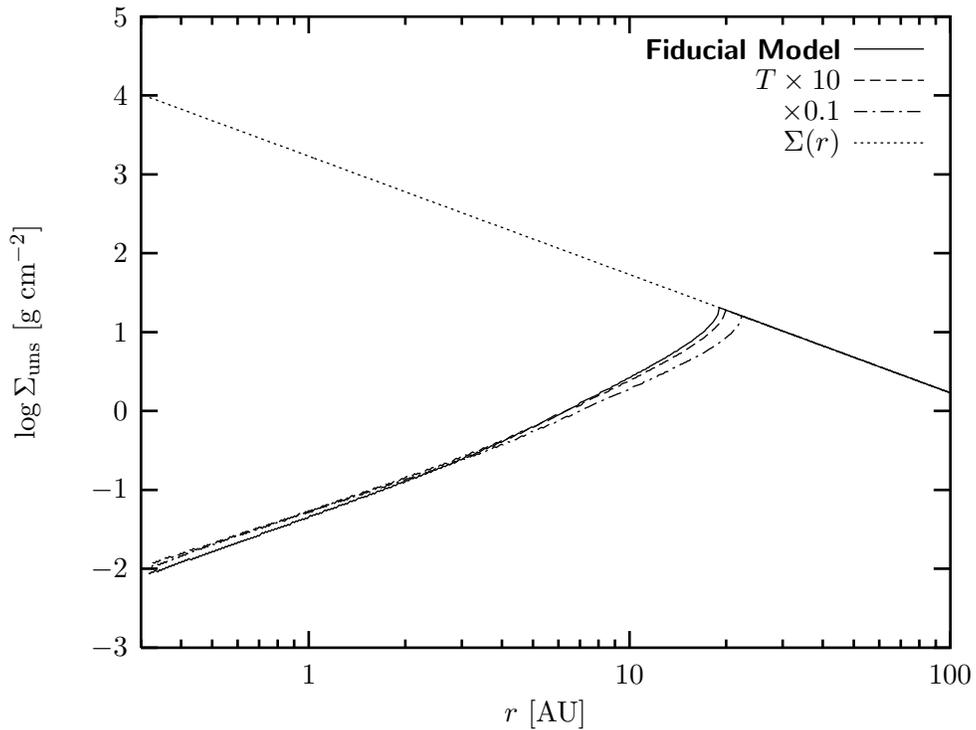
\begin{figure}
\begin{center}
\setlength{\unitlength}{0.1bp}
\begin{picture}(3600,2880)(0,0)
\put(3137,2304){\makebox(0,0)[r]{$\Sigma(r)$}}
\put(3137,2422){\makebox(0,0)[r]{$\times 0.1$}}
\put(3137,2540){\makebox(0,0)[r]{$T \times 10$}}
\put(3137,2658){\makebox(0,0)[r]{{\bsf Fiducial Model}}}
\put(2025,150){\makebox(0,0){$r$ [AU]}}
\put(100,1590){%
\makebox(0,0)[b]{\shortstack{$\log \Sigma_{\rm uns}$ [g~cm$^{-2}$]}}%
}
\put(3550,300){\makebox(0,0){$100$}}
\put(2341,300){\makebox(0,0){$10$}}
\put(1132,300){\makebox(0,0){$1$}}
\put(450,2780){\makebox(0,0)[r]{$5$}}
\put(450,2483){\makebox(0,0)[r]{$4$}}
\put(450,2185){\makebox(0,0)[r]{$3$}}
\put(450,1888){\makebox(0,0)[r]{$2$}}
\put(450,1590){\makebox(0,0)[r]{$1$}}
\put(450,1293){\makebox(0,0)[r]{$0$}}
\put(450,995){\makebox(0,0)[r]{$-1$}}
\put(450,698){\makebox(0,0)[r]{$-2$}}
\put(450,400){\makebox(0,0)[r]{$-3$}}
\end{picture} \\
\vspace{-12pt}
\caption
{The column density of the unstable layer, $\Sigma_{\rm uns}$, for the
models with temperature 10 times higher ({\it dashed curve}) and 10 times
lower ({\it dot-dashed curve}) than the fiducial model ({\it solid curve}).
The other parameters are the same as those of the fiducial model. 
\label{fig:temp}}
\end{center}
\end{figure}

The temperature distribution given by equation~(\ref{eqn:temp}) with $p
= 1 / 2$ holds for the disk transparent to the radiation from the
central star.
Actually, the inner dense regions of protoplanetary disks would be
optically thick and there would be some other heating sources such as
viscous heating and dissipation of magnetic fields.
However, the results are not very sensitive to the temperature as long
as it is below several hundred Kelvin where the thermal ionization is
inefficient. 
Figure~\ref{fig:temp} shows the column densities of the unstable layers
for the models with temperature ten times higher (dashed curve) and ten
times lower (dot-dashed curve) than the fiducial model (solid curve).
The other parameters are the same as those of the fiducial model.
The difference in $\Sigma_{\rm uns}$ is surprisingly small for such
large changes in the temperature.
We also investigated for $p = 0$ and $3 / 4$, and obtained quite
similar results assuming as above that the thermal ionization is
inefficient.
When the temperature is higher than several hundred Kelvin, the thermal
ionization becomes efficient and there would be significant effects on
the instability.
This would happen only in the innermost region $r \lesssim 0.1$ AU for
the minimum-mass model, which is much smaller than the dead zones
obtained in this paper.
Notice that the temperature of steady accretion disks ($p = 3 / 4$) with
typical mass accretion rate is within the range we examined here.

\subsection{Validity of Ionization-Recombination Equilibrium}

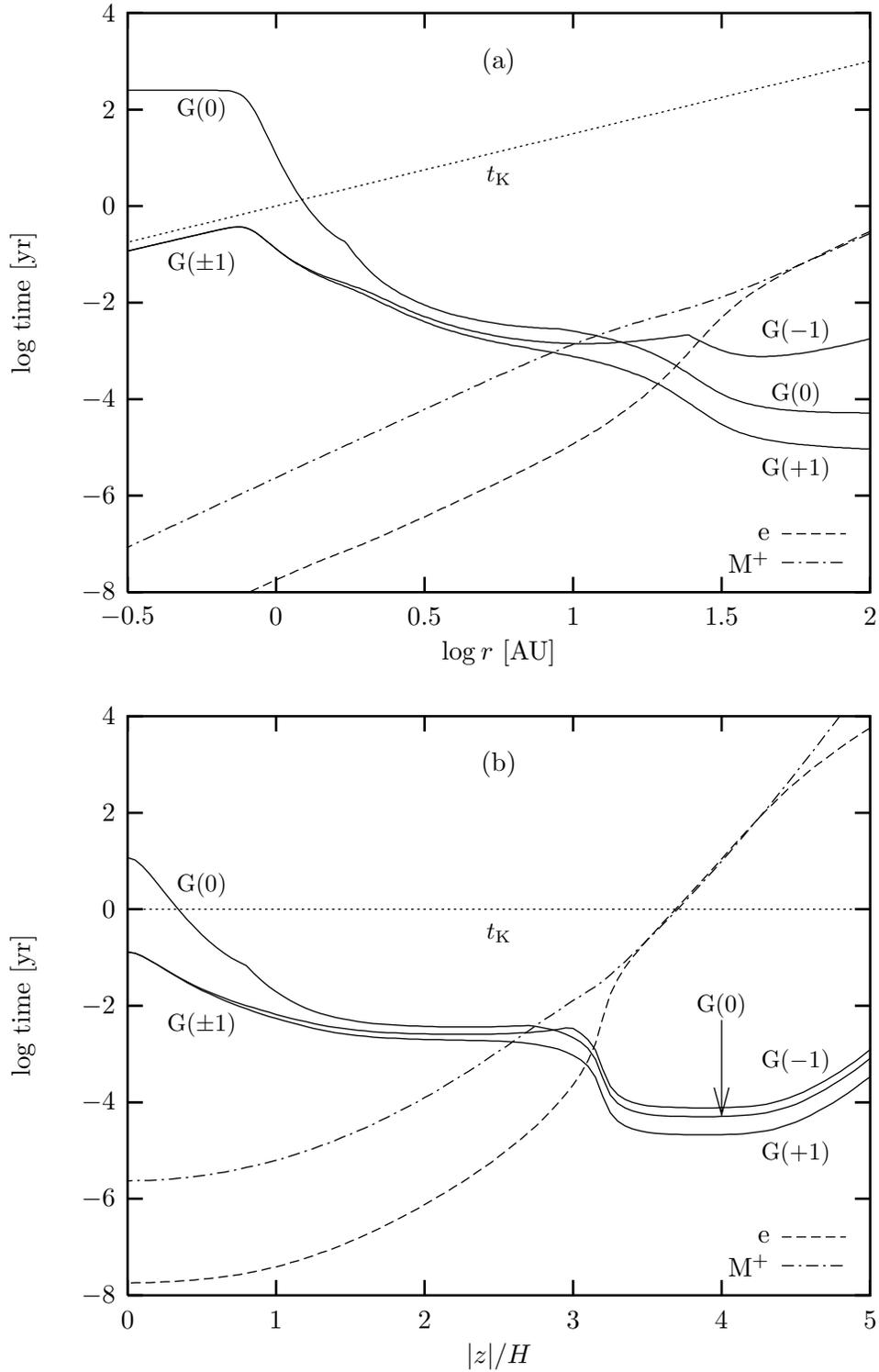
\begin{figure}
\begin{center}
\setlength{\unitlength}{0.1bp}
\begin{picture}(3600,2880)(0,0)
\put(3137,522){\makebox(0,0)[r]{M$^{+}$}}
\put(3137,640){\makebox(0,0)[r]{e}}
\put(2025,2582){\makebox(0,0){(a)}}
\put(2025,2111){\makebox(0,0){{\small $t_{\rm K}$}}}
\put(3245,1471){\makebox(0,0){{\small G($-$1)}}}
\put(3245,1213){\makebox(0,0){{\small G(0)}}}
\put(3245,906){\makebox(0,0){{\small G($+$1)}}}
\put(805,1749){\makebox(0,0){{\small G($\pm$1)}}}
\put(805,2383){\makebox(0,0){{\small G(0)}}}
\put(2025,150){\makebox(0,0){$\log r$ [AU]}}
\put(100,1590){%
\makebox(0,0)[b]{\shortstack{log time [yr]}}%
}
\put(3550,300){\makebox(0,0){$2$}}
\put(2940,300){\makebox(0,0){$1.5$}}
\put(2330,300){\makebox(0,0){$1$}}
\put(1720,300){\makebox(0,0){$0.5$}}
\put(1110,300){\makebox(0,0){$0$}}
\put(500,300){\makebox(0,0){$-0.5$}}
\put(450,2780){\makebox(0,0)[r]{$4$}}
\put(450,2383){\makebox(0,0)[r]{$2$}}
\put(450,1987){\makebox(0,0)[r]{$0$}}
\put(450,1590){\makebox(0,0)[r]{$-2$}}
\put(450,1193){\makebox(0,0)[r]{$-4$}}
\put(450,797){\makebox(0,0)[r]{$-6$}}
\put(450,400){\makebox(0,0)[r]{$-8$}}
\end{picture} \\
\setlength{\unitlength}{0.1bp}
\begin{picture}(3600,2880)(0,0)
\put(3137,522){\makebox(0,0)[r]{M$^{+}$}}
\put(3137,640){\makebox(0,0)[r]{e}}
\put(2025,2582){\makebox(0,0){(b)}}
\put(2025,1888){\makebox(0,0){{\small $t_{\rm K}$}}}
\put(3245,1362){\makebox(0,0){{\small G($-$1)}}}
\put(2940,1590){\makebox(0,0){{\small G(0)}}}
\put(3245,985){\makebox(0,0){{\small G($+$1)}}}
\put(805,1511){\makebox(0,0){{\small G($\pm$1)}}}
\put(805,2086){\makebox(0,0){{\small G(0)}}}
\put(2025,150){\makebox(0,0){$|z| / H$}}
\put(100,1590){%
\makebox(0,0)[b]{\shortstack{log time [yr]}}%
}
\put(3550,300){\makebox(0,0){$5$}}
\put(2940,300){\makebox(0,0){$4$}}
\put(2330,300){\makebox(0,0){$3$}}
\put(1720,300){\makebox(0,0){$2$}}
\put(1110,300){\makebox(0,0){$1$}}
\put(500,300){\makebox(0,0){$0$}}
\put(450,2780){\makebox(0,0)[r]{$4$}}
\put(450,2383){\makebox(0,0)[r]{$2$}}
\put(450,1987){\makebox(0,0)[r]{$0$}}
\put(450,1590){\makebox(0,0)[r]{$-2$}}
\put(450,1193){\makebox(0,0)[r]{$-4$}}
\put(450,797){\makebox(0,0)[r]{$-6$}}
\put(450,400){\makebox(0,0)[r]{$-8$}}
\end{picture} \\
\vspace{-12pt}
\caption
{The relaxation time $t_{\rm r}$ to the ionization-recombination
equilibrium for some representative particles (a) on the midplane as
functions of $r$, and (b) at $r = 1$ AU as functions of $|z|$, for the
fiducial model.
The Keplerian orbital period $t_{\rm K}$ is also shown by the dotted
line. 
\label{fig:ti}}
\end{center}
\end{figure}

We have used the densities of charged particles obtained by assuming
that the gas is in ionization-recombination equilibrium.
This assumption is valid when the relaxation time to an equilibrium
state for each kind of charged particles is much shorter than the
characteristic time scale of protoplanetary disks.
The relaxation time $t_{\rm r}({\rm X}_i)$ to the
ionization-recombination equilibrium for a species X$_{i}$ is
approximately given by the minimum of 
\begin{equation}
\left| \frac{n({\rm X}_i)}{\beta_{ijk} n({\rm X}_j)n({\rm X}_k)}
\right| \;.
\end{equation}
We estimate $t_{\rm r}({\rm X}_i)$ by using the equilibrium abundances
of particles adopted in the previous section.  
Figure~\ref{fig:ti}a shows the time scales near the midplane as
functions of $r$, and Figure~\ref{fig:ti}b shows those in the Earth's
region ($r = 1$ AU) as functions of $|z|$, for the fiducial model.

As seen from Figure~\ref{fig:ti}a, the gas near the midplane is almost
in ionization-recombination equilibrium at $r \gtrsim 1.3$ AU because
$t_{\rm r}$ is smaller than $t_{\rm K}$ for all the particles.  
At $r \lesssim 1$ AU, however, $t_{\rm r}$ for neutral grains is much
longer than $t_{\rm K}$.  
Because $t_{\rm r}$ for ions and electrons is much shorter than
$t_{\rm r}$ for grains, ions and electrons always take
quasi-equilibrium abundances determined for the temporary charge-state
distribution of grains.
When the charge-state distribution deviates significantly from the
equilibrium, the relaxation time to its true equilibrium is about 250
yr, much longer than $t_{\rm K}$ (see Fig~\ref{fig:ti}a). 
However, because this is much shorter than the sedimentation time of
grains in this region of the solar nebula (Nakagawa et al. 1981), the
ionization-recombination equilibrium can be attained even in this
region. 

In Earth's region the time scale $t_{\rm r}$ for metal ions and
electrons is longer than $t_{\rm K}$ only at $|z| \gtrsim 3.7 ~H$ as
seen from Figure~\ref{fig:ti}b.
Because most of the disk matter is inside this layer, a possible
deviation from the ionization-recombination equilibrium hardly affects
the results obtained in the previous sections and subsections. 

As the depletion factor $f_{\rm g}$ of dust grains decreases, the
time scale $t_{\rm r}$ for dust grains decreases considerably in
regions where ions and electrons are the dominant charged particles.  
For ions and electrons $t_{\rm r}$ remains at least 10 times smaller
than $t_{\rm K}$ even at such late evolutionary stages as $f_{\rm g}
= 10^{-4}$.
In the inner regions where dust grains are the dominant charged
particles, $t_{\rm r}$ for dust grains increases extensively as $r$
decreases, and it finally exceeds $t_{\rm K}$ at a certain position.  
For the case of $f_{\rm g} = 10^{-4}$ the time scale $t_{\rm r}$ for
charged grains exceeds $t_{\rm K}$ only at $r \lesssim 0.66$ AU, and
its value is at most 300 times longer than $t_{\rm K}$ even in the
innermost region $r \lesssim 0.5$ AU.  
Because the evolutionary time scale of the protoplanetary disk is much
longer than $t_{\rm K}$, we can regard that the solar nebula is
in ionization-recombination equilibrium even at stages of $f_{\rm g}
\approx 10^{-4}$ except at high $|z|$ where little matter exists. 
The situation is qualitatively the same for the other disks
investigated in the previous section.

\section{SUMMARY}

We have investigated the ionization state and magnetorotational
instability for various models of protoplanetary disks taking the effect
of charged grains into account. 
Our findings are summarized as follows.

\begin{enumerate}
\item 
In early evolutionary stages of the minimum-mass solar nebula, we have 
found that magnetorotational instability operates outside the critical 
radius $r_{\rm crit} \approx 20$ AU.
Inside the critical radius most part of the disk is stabilized by
ohmic dissipation.
Thus, the accreting matter accumulates near the critical radius unless
some other mechanisms of angular momentum transport are effective.

\item 
Just outside this critical radius, the mass accretion rate is $10^{-7}$
-- $10^{-6} ~M_{\odot}~{\rm yr}^{-1}$. 
By accumulation of matter at this rate the inner region becomes
gravitationally unstable within the typical lifetime of protoplanetary
disks.
This suggests that non-steady accretion onto the central star occurs as
in FU Orionis objects.

\item 
The critical radius $r_{\rm crit}$ depends on the mass of the disk.
Because the abundances of charged particles are lower in denser
regions, the disks of higher mass have larger stable regions and thus
larger $r_{\rm crit}$.

\item 
As dust grains in the disk grow or sediment toward the equatorial
plane, the critical radius $r_{\rm crit}$ decreases.
This suggests that magnetorotational instability is important in late
evolutionary stages of the disks exciting turbulence and suppressing
sedimentation of dust grains.
Therefore, to clarify the processes of planet formation, behavior of
grains should be studied in connection with gas motion.

\item 
Even in the inner regions $r \lesssim r_{\rm crit}$, some part of the
surface layer is magnetorotationally unstable.
However, because the column density of this layer is so low unless
the grain fraction is as low as $f_{\rm g} \lesssim 10^{-4}$ that the
layered accretion must not be effective at least in the early
evolutionary stages.
\end{enumerate}

\acknowledgments
Numerical computations were partly carried out at the Astronomical
Data Analysis Center of the National Astronomical Observatory, Japan.
This work was partly supported by Grant-in-Aid for Scientific Research 
on Priority Areas (10147105) of the Ministry of Education, Science,
Sports, and Culture of Japan.

\appendix
\section{MOMENTUM TRANSFER RATE BETWEEN ELECTRONS AND NEUTRAL PARTICLES}

M. Hayashi (1981) compiled the values of the momentum transfer cross
sections of an electron in collision with a hydrogen molecule and a
helium atom at low energies, which are not much different from the
geometrical cross sections and are significantly smaller than the
Langevin's values.
From these values we have obtained the empirical formulae for the
momentum transfer rate coefficients as functions of the temperature
$T$ as follows,
\begin{equation}
\langle \sigma v \rangle_{{\rm e-H}^2} \approx
\left( \frac{8 k T}{\pi m} \right)^{1/2}
\left[  7.94 \times 10^{-16} 
- 2.52 \times 10^{-16} \log T
+ 1.19 \times 10^{-16} ( \log T )^2
\right] \;,
\label{eqn:h2}
\end{equation}
\begin{equation}
\langle \sigma v \rangle_{\rm e-He} \approx
\left( \frac{8 k T}{\pi m} \right)^{1/2}
\left[ 5.30 \times 10^{-16} 
- 4.70 \times 10^{-16} \log T
+ 2.31 \times 10^{-16} ( \log T )^2
\right] \;,
\label{eqn:he}
\end{equation}
where $m$ is the reduced mass.

Equations~(\ref{eqn:h2}) and (\ref{eqn:he}) give the rate coefficients 
about two orders of magnitude smaller than those used previously by
many authors.
Thus, we find that electrons are much more tightly coupled with
magnetic field than previously considered. 


\clearpage

\clearpage

\clearpage

\clearpage

\clearpage

\begin{figure}
\begin{center}
\setlength{\unitlength}{0.1bp}
\begin{picture}(3600,1080)(0,0)
\put(100,615){%
\makebox(0,0)[b]{\shortstack{$\log \zeta ~[ {\rm s}^{-1} ]$}}%
}
\put(500,858){\makebox(0,0)[r]{$-17$}}
\put(500,696){\makebox(0,0)[r]{$-19$}}
\put(500,534){\makebox(0,0)[r]{$-21$}}
\put(500,372){\makebox(0,0)[r]{$-23$}}
\end{picture} \\
\vspace{-36pt}
\setlength{\unitlength}{0.1bp}
\begin{picture}(3600,2880)(0,0)
\put(1100,2294){\makebox(0,0)[r]{H$_3^{+}$}}
\put(1100,2412){\makebox(0,0)[r]{m$^{+}$}}
\put(1100,2530){\makebox(0,0)[r]{M$^{+}$}}
\put(1100,2648){\makebox(0,0)[r]{e}}
\put(3175,1888){\makebox(0,0){{\small G($-$2)}}}
\put(3175,1656){\makebox(0,0){{\small G($-$1)}}}
\put(3175,1412){\makebox(0,0){{\small G(0)}}}
\put(3175,684){\makebox(0,0){{\small G($+$2)}}}
\put(3175,1041){\makebox(0,0){{\small G($+$1)}}}
\put(2800,2463){\makebox(0,0){{\small e, M$^{+}$}}}
\put(2050,150){\makebox(0,0){$|z| / H$}}
\put(100,1590){%
\makebox(0,0)[b]{\shortstack{$\log [ n({\rm X}) / n_{\rm H} ]$}}%
}
\put(3550,300){\makebox(0,0){$4$}}
\put(2800,300){\makebox(0,0){$3$}}
\put(2050,300){\makebox(0,0){$2$}}
\put(1300,300){\makebox(0,0){$1$}}
\put(550,300){\makebox(0,0){$0$}}
\put(500,2648){\makebox(0,0)[r]{$-6$}}
\put(500,2383){\makebox(0,0)[r]{$-8$}}
\put(500,2119){\makebox(0,0)[r]{$-10$}}
\put(500,1854){\makebox(0,0)[r]{$-12$}}
\put(500,1590){\makebox(0,0)[r]{$-14$}}
\put(500,1326){\makebox(0,0)[r]{$-16$}}
\put(500,1061){\makebox(0,0)[r]{$-18$}}
\put(500,797){\makebox(0,0)[r]{$-20$}}
\put(500,532){\makebox(0,0)[r]{$-22$}}
\end{picture} \\
\vspace{-12pt}
\caption
{Relative abundances $n({\rm X})/n_{\rm H}$ of some representative
particles as functions of the height $|z|$ from the midplane at $r =
30$ AU for the same disk as in Figure~{\protect{\ref{fig:rdep}}}.
The upper panel shows the ionization rate $\zeta$.
\label{fig:zdep2}}
\end{center}
\end{figure}

\clearpage

\begin{figure}
\begin{center}
\setlength{\unitlength}{0.1bp}
\begin{picture}(3600,2880)(0,0)
\put(3137,522){\makebox(0,0)[r]{$r = 1$ AU ($f_{\rm g} = 10^{-4}$)}}
\put(3137,640){\makebox(0,0)[r]{$r = 30$ AU ($f_{\rm g} = 1$)}}
\put(3137,758){\makebox(0,0)[r]{$r = 1$ AU ($f_{\rm g} = 1$)}}
\put(2050,150){\makebox(0,0){$|z| / H$}}
\put(100,1590){%
\makebox(0,0)[b]{\shortstack{$\log ( t_{\rm dis} / t_{\rm K} )$}}%
}
\put(3550,300){\makebox(0,0){$5$}}
\put(2950,300){\makebox(0,0){$4$}}
\put(2350,300){\makebox(0,0){$3$}}
\put(1750,300){\makebox(0,0){$2$}}
\put(1150,300){\makebox(0,0){$1$}}
\put(550,300){\makebox(0,0){$0$}}
\put(500,2423){\makebox(0,0)[r]{$5$}}
\put(500,1828){\makebox(0,0)[r]{$0$}}
\put(500,1233){\makebox(0,0)[r]{$-5$}}
\put(500,638){\makebox(0,0)[r]{$-10$}}
\end{picture} \\
\vspace{-12pt}
\caption
{The ratio of the ohmic dissipation time $t_{\rm dis}$ to the
Keplerian orbital period $t_{\rm K}$ as a function of $|z|$ at $r = 1$
AU ({\it solid curve}) and 30 AU ({\it dashed curve}).
The magnetic diffusivity is evaluated with the abundances of charged
particles shown in Figures~{\protect{\ref{fig:zdep}}} and
{\protect{\ref{fig:zdep2}}} for the fiducial model.
The dot-dashed curve is for the case of $f_{\rm g} = 10^{-4}$ at $r = 1$
AU.
\label{fig:ztd}}
\end{center}
\end{figure}
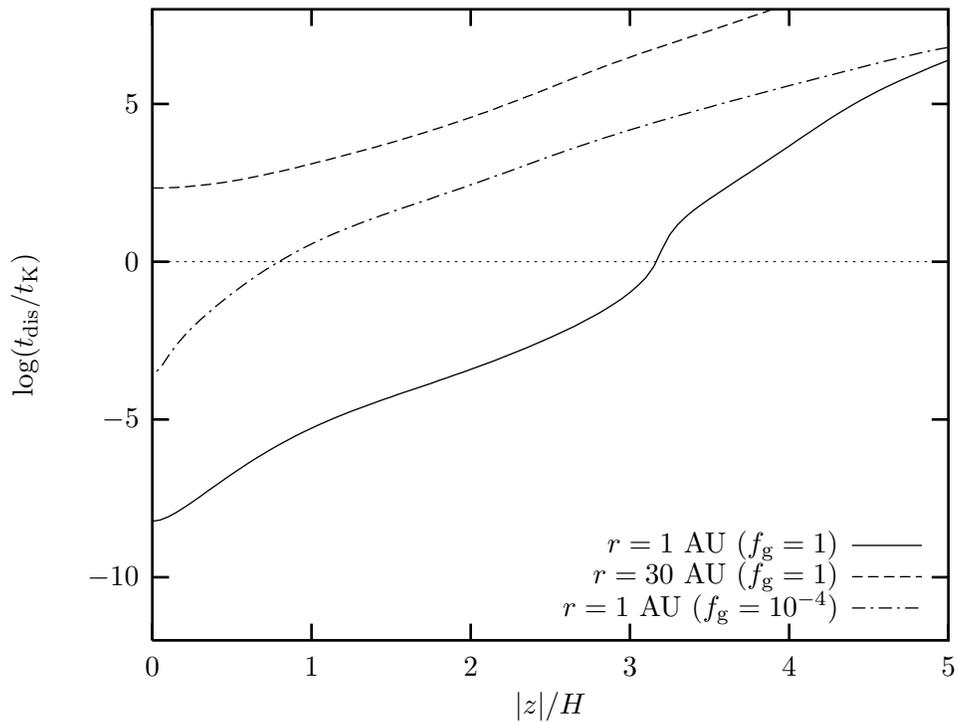

\clearpage

\clearpage

\clearpage

\clearpage

\clearpage

\clearpage

\clearpage

\clearpage

\clearpage

\clearpage

\clearpage 

\begin{deluxetable}{clcc}
\small
\tablewidth{0pt}
\tablecaption{Abundance of Elements and their Fraction in the Gas Phase
\label{tbl:ele}}
\tablehead{
\colhead{Element} & 
\colhead{Abundance} & 
\colhead{Chemical Species} & 
\colhead{Fraction in Gas Phase}}
\startdata
H & 1 & H$_2$ & 1 \nl
He & $9.75 \times 10^{-2}$ & He & 1 \nl
C & $3.62 \times 10^{-4}$ & CO & $\delta_1$ \nl
O & $8.53 \times 10^{-4}$ & O$_2$,O & $\delta_1$ \nl
Metal Total & $7.97 \times 10^{-5}$ & M & $\delta_2$ 
\enddata
\end{deluxetable}

\begin{deluxetable}{lclclclcl@{\hspace{30pt}}l} 
\small
\tablewidth{0pt}
\tablecolumns{10}
\tablecaption{Two-Body Reactions and their Rate Coefficients
\label{tbl:two}}
\tablehead{
\multicolumn{7}{c}{Reaction} & \colhead{} & \colhead{} &
\colhead{Rate coefficient (cm$^3$ s$^{-1}$)}}
\startdata
H$^+$ & + & Mg & $\rightarrow$ & Mg$^+$ & + & H & & & 
$1.10 \times 10^{-9}$ \nl
H$^+$ & + & O & $\rightarrow$ & O$^+$ & + & H & & & 
$7.00 \times 10^{-10} \exp ( - 232.0 / T )$ \nl
H$^+$ & + & O$_2$ & $\rightarrow$ & O$_2^+$ & + & H & & & 
$2.00 \times 10^{-9}$ \nl
H$_2^+$ & + & H$_2$ & $\rightarrow$ & H$_3^+$ & + & H & & &
$2.08 \times 10^{-9}$ \nl
H$_3^+$ & + & O & $\rightarrow$ & OH$^+$ & + & H$_2$ & & & 
$8.00 \times 10^{-10}$ \nl
H$_3^+$ & + & Mg & $\rightarrow$ & Mg$^+$ & + & H & + & H$_2$ & 
$1.00 \times 10^{-9}$ \nl
H$_3^+$ & + & CO & $\rightarrow$ & HCO$^+$ & + & H$_2$ & & & 
$1.70 \times 10^{-9}$ \nl
H$_3^+$ & + & O$_2$ & $\rightarrow$ & O$_2$H$^+$ & + & H$_2$ & & & 
$5.00 \times 10^{-9} \,\, \exp ( - 150.0 / T )$ \nl
He$^+$ & + & H$_2$ & $\rightarrow$ & H$^+$ & + & H & + & He & 
$3.70 \times 10^{-14} \exp ( - 35.0 / T)$ \nl
He$^+$ & + & CO & $\rightarrow$ & C$^+$ & + & O & + & He & 
$1.60 \times 10^{-9}$ \nl
He$^+$ & + & O$_2$ & $\rightarrow$ & O$^+$ & + & O & + & He & 
$1.00 \times 10^{-9}$ \nl
C$^+$ & + & Mg & $\rightarrow$ & Mg$^+$ & + & C & & & 
$1.10 \times 10^{-9}$ \nl
C$^+$ & + & H$_2$ & $\rightarrow$ & CH$_2^+$ & + & $h \nu$ & & & 
$4.00 \times 10^{-16} ( T / 300 )^{-0.20}$ \nl
C$^+$ & + & O$_2$ & $\rightarrow$ & CO$^+$ & + & O & & & 
$3.80 \times 10^{-10}$ \nl
C$^+$ & + & O$_2$ & $\rightarrow$ & O$^+$ & + & CO & & & 
$6.20 \times 10^{-10}$ \nl
HCO$^+$ & + & Mg & $\rightarrow$ & Mg$^+$ & + & HCO & & & 
$2.90 \times 10^{-9}$ \nl \nl
H$^+$ & + & e & $\rightarrow$ & H & + & $h \nu$ & & & 
$3.50 \times 10^{-12} ( T / 300 )^{-0.75}$ \nl
H$_3^+$ & + & e & $\rightarrow$ & H$_2$ & + & H & & & 
$1.50 \times 10^{-8} \,\, ( T / 300 )^{-0.50}$ \nl
H$_3^+$ & + & e & $\rightarrow$ & H & + & H & + & H & 
$1.50 \times 10^{-8} \,\, ( T / 300 )^{-0.50}$ \nl
He$^+$ & + & e & $\rightarrow$ & He & + & $h \nu$ & & & 
$2.36 \times 10^{-12} ( T / 300 )^{-0.64}$ \nl
C$^+$ & + & e & $\rightarrow$ & C & + & $h \nu$ & & & 
$4.40 \times 10^{-12} ( T / 300 )^{-0.61}$ \nl
Mg$^+$ & + & e & $\rightarrow$ & Mg & + & $h \nu$ & & & 
$2.80 \times 10^{-12} ( T / 300 )^{-0.86}$ \nl
HCO$^+$ & + & e & $\rightarrow$ & CO & + & H & & & 
$1.10 \times 10^{-7} \,\, ( T / 300 )^{-1.00}$ 
\enddata
\end{deluxetable}

\begin{deluxetable}{lclclcl@{\hspace{30pt}}l} 
\small
\tablewidth{0pt}
\tablecaption{Ionization Rates \label{tbl:ionize}}
\tablecolumns{8}
\tablehead{
\multicolumn{3}{c}{Reaction} & \colhead{} & \colhead{} & \colhead{} &
\colhead{} & \colhead{Rate}}
\startdata
H$_2$ & $\rightarrow$ & H$_2^+$ & + & e & & & $0.97 \zeta$\tablenotemark{a} \nl
H$_2$ & $\rightarrow$ & H$^+$ & + & H & + & e & $0.03 \zeta$\nl
He & $\rightarrow$ & He$^+$ & + & e & & & $0.84 \zeta$
\enddata
\tablenotetext{a}{$\zeta$ is the total
ionization rate for a hydrogen molecule.}
\end{deluxetable}

\end{document}